\documentclass{aa}  
\usepackage{graphicx}
\usepackage{txfonts}
\usepackage{natbib}
\usepackage{lscape}
\usepackage{appendix}

\newcommand{\hii}{H\,{\small II}\xspace}
\newcommand{\hi}{H\,{\small I}\xspace}

\begin{document} 

\title{Radio and infrared study of southern \hii\ regions G346.056$-$0.021 and G346.077$-$0.056}

\author {Swagat R Das\inst{1}, Anandmayee. Tej\inst{1}, Sarita Vig\inst{1}, Tie Liu \inst{2,3},
   Swarna K. Ghosh\inst{4}, Ishwara Chandra C.H.\inst{4}}

\institute{Indian Institute of Space Science and Technology,
              Trivandrum 695547, India\\
              \email{swagat.12@iist.ac.in}
              \and 
  Korea Astronomy and Space Science Institute 776, Daedeokdae-ro,
Yuseong-gu, Daejeon, Republic of Korea 305-348 
\and
East Asian Observatory, 660 N. A'ohoku Place, Hilo, HI 96720, USA 
         \and
             National Centre For Radio Astrophysics, Pune 411007, India\\}


 
  \abstract
   {}
   {We present a multiwavelength study of two southern Galactic \hii\ regions 
G346.056$-$0.021 and G346.077$-$0.056 which are located at a distance of 10.9~kpc. 
The distribution of ionized gas, cold and warm dust and the stellar population
associated with the two \hii\ regions are studied in detail using
measurements at near-infrared, mid-infrared, far-infrared, submillimeter and radio wavelengths.}
   {The radio continuum maps at 1280 and 610 MHz were obtained using the Giant Metrewave Radio Telescope to probe the ionized gas. The dust temperature, column density and dust emissivity maps were generated by using modified blackbody fits in the far-infrared wavelength range 160 - 500~$\rm \mu m$. Various near- and mid-infrared colour and magnitude criteria were adopted
   to identify candidate ionizing star(s) and the population of young stellar objects in the associated field. 
   
   }
   {The radio maps reveal the presence diffuse ionized emission  displaying
   distinct cometary morphologies. The 1280~MHz flux densities translate to ZAMS 
   spectral types in the range O7.5V - O7V and O8.5V - O8V for the ionizing
   stars of G346.056$-$0.021 and G346.077$-$0.056, respectively.  A few
   promising candidate ionizing star(s) are identified using near-infrared
   photometric data. The column 
   density map shows the presence of a large, dense dust clump enveloping G346.077$-$0.056. The dust temperature map shows peaks towards the two \hii\ regions.
   The submillimetre image shows the presence of two additional clumps one being
   associated with G346.056$-$0.021. The masses of the clumps are estimated to range between 
   $\sim$ 1400 to 15250 $\rm M_{\odot}$. Based on simple analytic calculations and
   the correlation seen between the ionized gas distribution and the local density structure, the observed 
   cometary morphology in the radio maps is better explained invoking the champagne-flow model.     }


\authorrunning{Das et al.}
\titlerunning{Multi-wavelength study of HII regions G346.056$-$0.021 and G346.077$-$0.056}

\maketitle
%
\section{Introduction} \label{intro}
Massive O and B star formation is accompanied by enormous Lyman continuum emission.
The outpouring of UV photons ionize the surrounding interstellar medium (ISM) forming 
\hii\ regions and thus revealing the location of ongoing high-mass star formation 
through radio free-free emission. In the evolutionary sequence, it starts with the
deeply embedded, hypercompact \hii\ regions which eventually expands forming the ultra-compact (UC \hii\ ), compact and extended or classical \hii\ regions. The earliest evolutionary phase is 
closely linked to the formation process where the newly born massive star is still in the 
accretion phase. 
The classical \hii\ regions, on the other hand, are mostly associated with more evolved objects. Apart
from being bright in the radio, the high-luminosity of the massive stars also makes the \hii\ 
regions bright in the infrared (IR). The UV and optical
radiation from the star is absorbed by the dust and is re-emitted in the IR.The radio and the thermal IR being least affected by extinction, studies in these
wavelength regimes 
allow us to probe deep into the star forming clouds to unravel the processes associated with high-mass star formation and the cooler dust environment. Complimentary 
near-infrared (NIR) studies further provides the census of the associated young stellar population.  Excellent reviews on the nature and physical properties
of \hii\ regions can be found in \citet{{2007prpl.conf..181H},{1989ApJS...69..831W},{2002ARA&A..40...27C},{1999PASP..111.1049G}}.
\begin{figure*}
\centering
\includegraphics[scale=0.4]{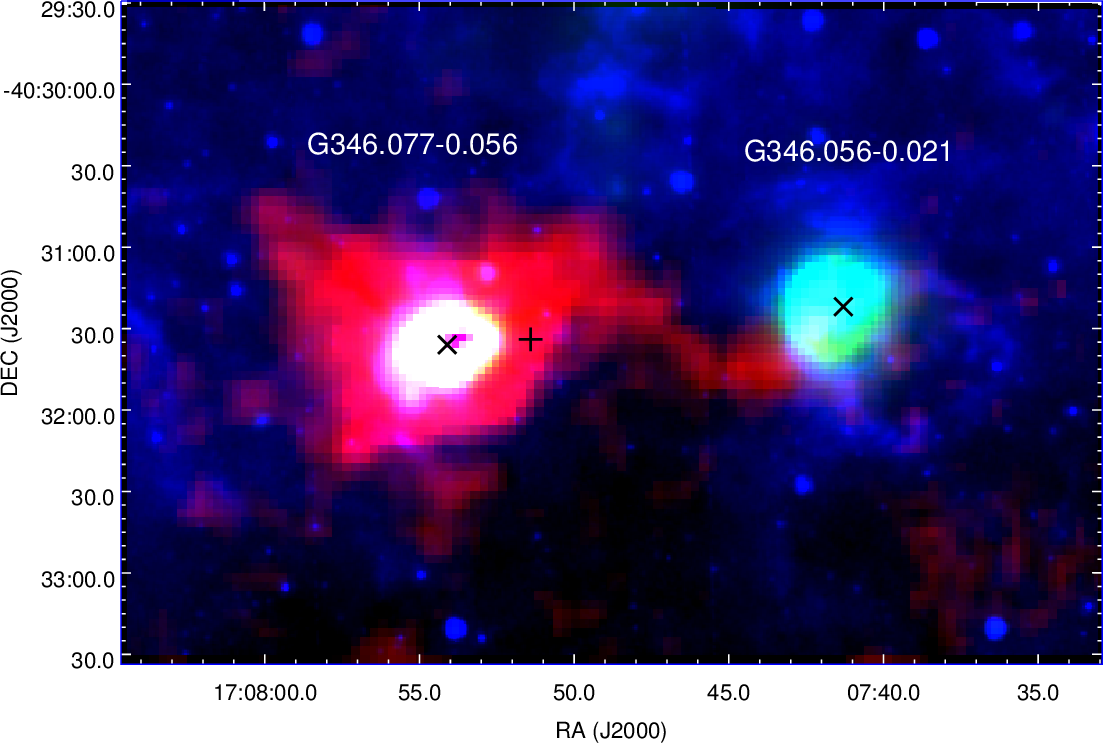}
\caption {Colour composite image of the region associated with G346.056$-$0.021 and G346.077$-$0.056 
using IRAC
8.0~$\rm \mu m$ band (see section \ref{spitzer_archive}) (blue), MIPS 24~$\rm \mu m$  (MIPSGAL Survey; \citealt{2009PASP..121...76C}) (green), and ATLASGAL
870~$\rm \mu m$ (see section \ref{ATLASGAL})
(red). Positions of the \hii\ regions as listed in \citet{2011ApJS..194...32A} are shown with the 
`$\times$' symbols. The + mark shows the position of IRAS 17043$-$4027. It should be noted here 
that a few pixels towards the central part of the 24~$\rm \mu m$ emission associated with 
G346.077$-$0.056 are saturated.} 
\label{region_8_24_850}
\end{figure*}

In this paper, we study two southern \hii\ regions from IR through radio wavelengths. 
These are G346.056$-$0.021 and G346.077$-$0.056. IRAS 17043$-$4027, 
with IRAS colours consistent with UC \hii\ regions \citep{1996A&AS..115...81B}, is associated with G346.077$-$0.056.
From the $\rm ^{13}CO$ observations of candidate massive young stellar objects (YSOs) in the southern Galactic plane,
\citet{2007A&A...474..891U} give the kinematic distance estimate of 5.7~kpc (near) and 10.8~kpc 
(far) for G346.056$-$0.021 and 5.8~kpc (near) and 10.7~kpc (far) for G346.077$-$0.056. In a later paper \citep{2014MNRAS.437.1791U}, they resolve the kinematic distance ambiguity 
by using \hi\ self-absorption analysis and place both the sources at a far distance of 10.9~kpc.  
We adopt this distance in our study.
As part of the Green Bank Telescope (GBT) \hii\ Region Discovery Survey, 
hydrogen radio recombination line (RRL) emission was detected
towards G346.056$-$0.021 and G346.077$-$0.056 \citep{2011ApJS..194...32A} and helium and carbon RRLs were
observed towards G346.056$-$0.021 \citep{2013ApJ...764...34W}. From their 1400~MHz 
Galactic plane survey, \citet{1990ApJS...74..181Z} have listed G346.056$-$0.021 and G346.077$-$0.056 as 
small diameter radio sources with sizes of  18\arcsec.8 and 9\arcsec, respectively. Apart from this, 
G346.077$-$0.056 was observed at 4800 and 8640~MHz by \citet{2007A&A...461...11U} as part of the Red MSX Sources 
(RMS) survey to obtain radio observations of candidate massive YSOs. Other than the $\rm ^{13}CO$ survey mentioned earlier, a few other 
molecular line studies have been conducted towards G346.077$-$0.056 \citep{{1996A&AS..115...81B},
{2013ApJ...777..157H},{2015MNRAS.446.2566Y}}. G346.077$-$0.056 has been part of a 6.7~GHz methanol maser 
detection survey which yielded negative results \citep{1995A&AS..110...81V}.
 Similar result was also obtained from the recent 6.7~GHz methanol maser survey \citep{2010MNRAS.404.1029C}.
 Further, a sparsely populated embedded cluster, VVV CL094, 
is reported to be associated 
with G346.077$-$0.056 \citep{{2011A&A...532A.131B},{2013A&A...560A..76M}}. As is evident
from the literature survey, there are no dedicated study on either of these
two \hii\ regions. 

Figure \ref{region_8_24_850} displays a colour-composite image of the  associated region showing the location and the dust environment. The 8~$\rm \mu m$ emission is relatively localized towards
the centre of G346.077$-$0.056 as compared to G346.056$-$0.021 where it displays an extended bubble
type morphology towards the south-west. 24~$\rm \mu m$ emission probing the warm dust component is seen predominantly
towards G346.077$-$0.056 and the central portion of G346.077$-$0.056. The 870~$\rm \mu m$ component
which traces the cold dust environment is seen as a prominent, large clump enveloping 
G346.077$-$0.056 with a distinct filamentary structure towards the west connecting to G346.056$-$0.021.  

In presenting a multiwavelength study of the complex associated with these two \hii\ regions, we have organized the paper in the following way. Section \ref{obs} describes the radio
continuum observation and the associated data reduction procedure. In this section, we have also discussed the various archival data used in this study. Study of the
ionized and the dust component and the associated stellar population is presented
in Section \ref{results}. Possible mechanisms responsible for the morphology of
the ionized emission for both \hii\ regions are explored in Section \ref{morph}.
The summary of the results obtained are compiled in Section \ref{summ}. 

\section{Observation and archival data sets}\label{obs}
\subsection{Radio continuum observations }\label{radio_obs} 
To probe the ionized emission associated with the \hii\ regions, radio continuum mapping were 
carried out at 610 and 1280~MHz using the Giant Metrewave Radio Telescope (GMRT), Pune India. 
GMRT contains 30 dishes of 45~m diameter each arranged in a hybrid `Y' shaped configuration. 
This ensures a wide UV coverage. The central square of GMRT has 12 antennae randomly arranged 
within a compact area of $\rm 1 \times 1~ km^2$. The remaining 18 antennae are placed in the 
arms of the `Y' with each arm comprising of six antennae. With the possible baselines ranging 
between $\sim$ 100~m and  $\sim$ 25~km, GMRT enables us to probe the ionized emission at
various resolutions and spatial scales. \citet{1991CuSc...60...95S} provides detail information 
regarding GMRT and its configuration. 

The continuum observations were carried out at 610 and 
1280~MHz with a bandwidth of 32~MHz. This was done in the spectral line mode to 
minimize the effects of bandwidth smearing and narrowband RFI. The on-source integration time is $\sim 4$ hr.
The radio sources 3C48 and 3C286 were used as primary flux calibrators and 1626$-$298 was used as a
phase calibrator. These provide the amplitude and phase gains for flux and phase calibration of
the measured visibilities respectively. Data reduction was carried out using Astronomical Image Processing 
System (AIPS). The tasks {\tt UVPLT}, {\tt VPLOT} and {\tt TVFLG} were used to check the data 
carefully and bad data (dead antenna, dead baseline, spikes, RFI, etc) were edited out using the 
tasks {\tt TVFLG} and {\tt UVFLG}. To keep the bandwidth smearing effect negligible the calibrated 
data was averaged in frequency.  
We adopt the wide-field imaging technique to account for $w$-term effect. Several iterations of `phase-only' self calibration were done to minimize amplitude 
and phase errors and obtain better {\it rms} noise in the maps. Primary beam correction was 
applied with the task {\tt PBCOR}. 

While observing towards the Galactic plane, the contribution from the Galactic diffuse emission is appreciable and leads to the rise in system temperature. The flux calibration is based
on the sources which are off the Galactic plane. Hence, it becomes essential to rescale the
generated maps. G346.056$-$0.021 and G346.077$-$0.056 are located north-east of the bubble S10 which is studied in
detail in \citet{2016AJ....152..152R}. Both these \hii\ regions were observed
as part of the same field with S10 being at the phase centre. Thus we use the same
scaling factors of 1.2 (1280~MHz) and 1.7 (610~MHz) derived in \citet{2016AJ....152..152R}. 
 
\subsection{Archival data sets } \label{archive}

\subsubsection{Near-infrared data from VVV and 2MASS} \label{2mass_archive}

NIR ($\rm JHK/K_s)$) photometric data for point sources around our region of interest
are obtained from the VISTA Variables in Via Lactea (VVV; \citealt{2010NewA...15..433M}) and 
Two Micron All Sky Survey Point Source Catalog (2MASSPSC; \citealt{2006AJ....131.1163S}). The VVV and
2MASS images have a resolution of $\sim$~0.8$\arcsec$ and 5$\arcsec$, respectively. Good
quality photometric data are retrieved from these catalogs and used to study the nature of the stellar population associated with the two \hii\ regions.

\subsubsection{Mid-infrared data from Spitzer} \label{spitzer_archive}
Mid-infrared (MIR) data enclosing the two \hii\ regions have been obtained from the archive of
{\it Spitzer} Space Telescope. The Infra-Red Array Camera (IRAC) and Multiband Imaging Photometer 
(MIPS) are the two onboard instruments. Simultaneous images at 3.6, 4.5, 5.8, 8.0~$\rm \mu m$ is obtained by IRAC with angular resolutions $ < 2\arcsec$ \citep{2004ApJS..154...10F}. 
The Level-2 Post-Basic Calibrated Data (PBCD) images from the Galactic Legacy Infrared Mid-Plane Survey Extraordinaire (GLIMPSE ; \citealt{2003PASP..115..953B}) and photometric data from the
`highly reliable', GLIMPSE I Spring'07 catalog are
used in this paper. These data are used to study the 
population of young stellar objects (YSOs) and warm dust 
associated with the regions. 

\subsubsection{Far-infrared data from Herschel} \label{Herschel}
Far-infrared (FIR) maps for our regions in the wavelength range $\rm 70 - 500~\mu m$
have been retrieved from  the {\it Herschel} Space Observatory archives. These regions
were observed as part of the Herschel Infrared Galactic Plane Survey (HI-GAL; 
\citealt{2010A&A...518L.100M}). We have used the images obtained with the Photodetector Array 
Camera and Spectrometer (PACS; \citealt{2010A&A...518L...2P}) and Spectral and Photometric Imaging 
Receiver (SPIRE; \citealt{2010A&A...518L...3G}). We have used Level-2 PACS images at 70 and 
160~$\rm \mu m$  and Level-3 SPIRE images at 250, 350 and 500~$\rm \mu m$ images from the archive 
for our study. 
The images used have resolution of 
5\arcsec.9, 11\arcsec.6, 18\arcsec.5, 25\arcsec.3 and 36\arcsec.9 and pixel sizes of 3\arcsec.2, 
3\arcsec.2, 6\arcsec, 10\arcsec~ and 14\arcsec~ 
at 70, 160, 250, 350, and 500~$\rm \mu m$, respectively. The {\it Herschel} Interactive
Processing Environment (HIPE)\footnote{HIPE is a joint development by the Herschel Science Ground 
Segment Consortium, consisting of ESA, the NASA Herschel Science Center, and the HIFI, PACS and 
SPIRE consortia.} is used for processing the images. 
We have used the FIR data to study the physical properties of cold dust emission associated with 
the regions.

\subsubsection{870~$\rm \mu m$ data from ATLASGAL} \label{ATLASGAL}
The 870~$\rm \mu m$ image used in this study has been obtained from the archives of the Apex 
Telescope under the APEX Telescope Large Area Survey of the Galaxy (ATLASGAL)\footnote{This project is a 
collaboration between the Max Planck Gesellschaft (MPG: Max Planck Institute für Radioastronomie, MPIfR Bonn, 
and Max Planck Institute for Astronomie, MPIA Heidelberg), the European Southern Observatory (ESO) and the 
Universidad de Chile} which used the LABOCA bolometer array \citep{2009A&A...504..415S}. The resolution of 
ATLASGAL image is 18.2$\arcsec$. This data is used to study the properties of cold dust clumps associated with 
these \hii\ regions.

\section{Results and Discussion} \label{results}

\subsection{Emission from ionized gas} \label{ionized}
\begin{figure*}
\centering
\includegraphics[scale=0.25]{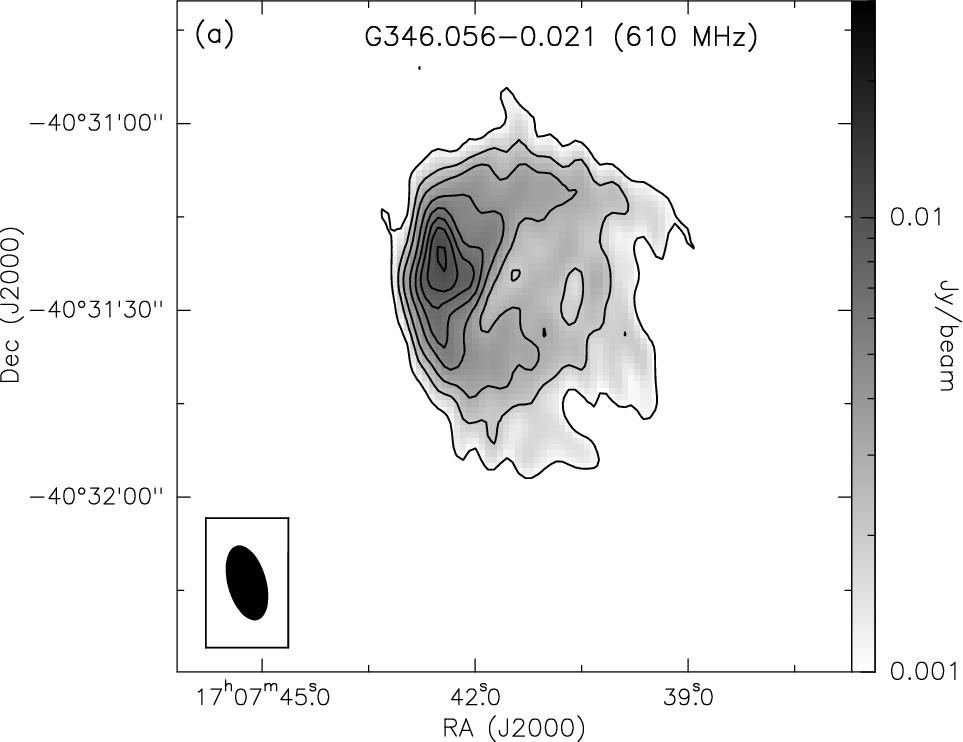}
\includegraphics[scale=0.25]{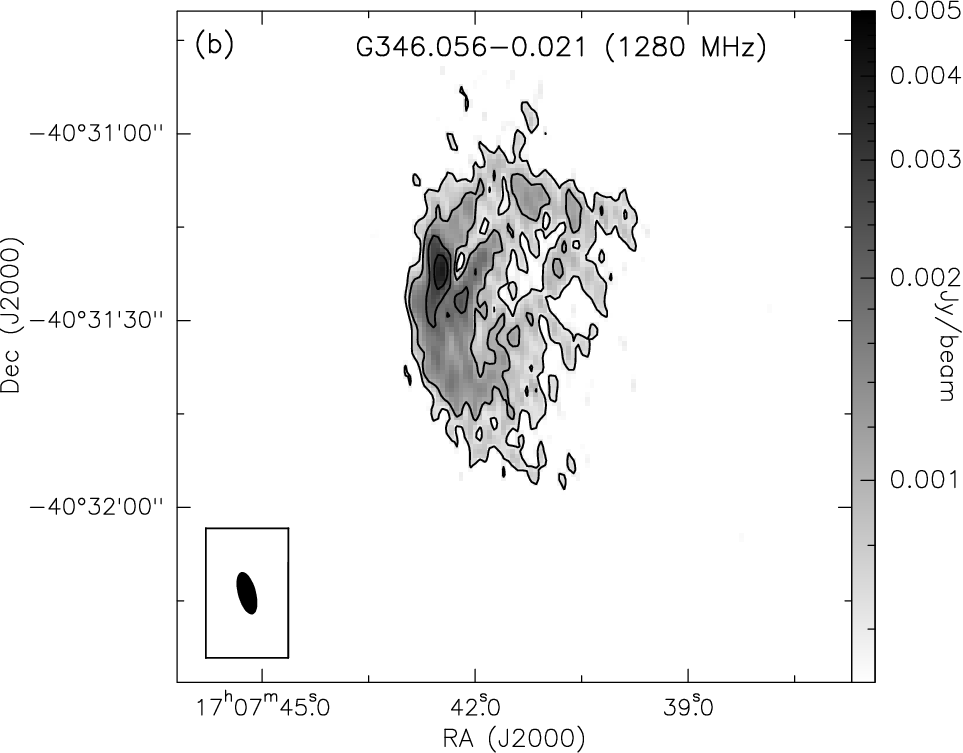}
\includegraphics[scale=0.25]{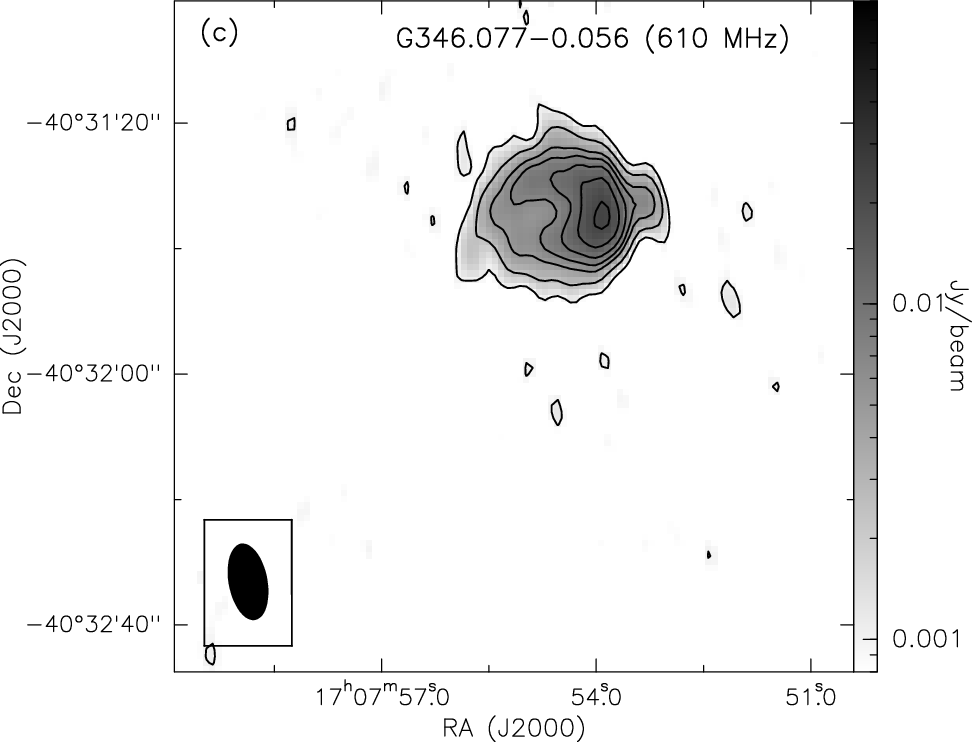}
\includegraphics[scale=0.25]{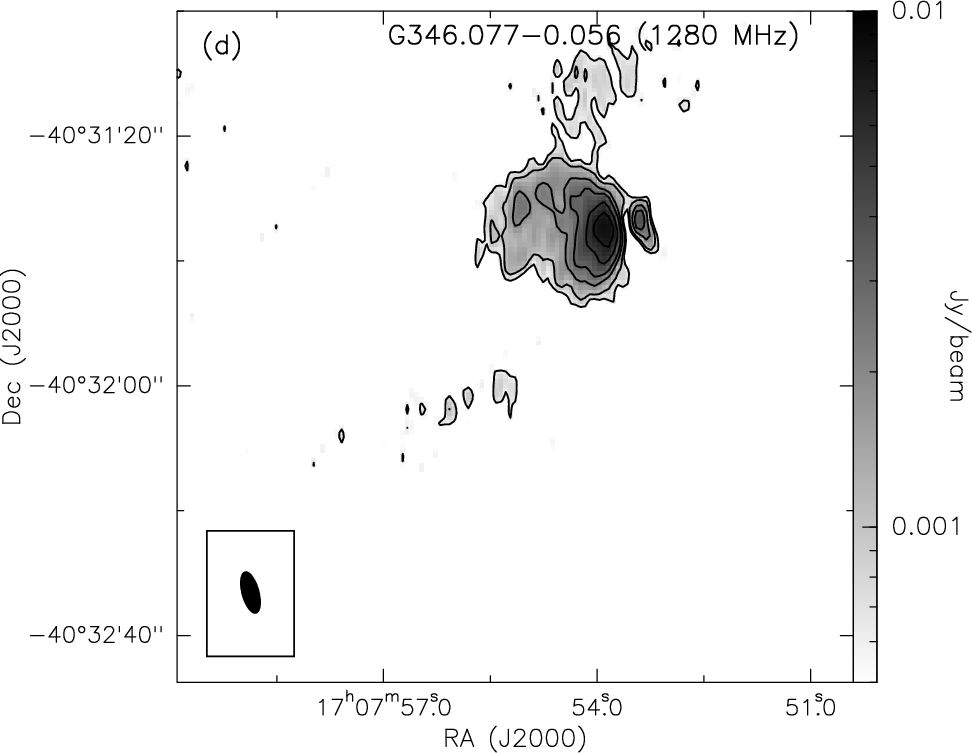}
\caption {Ionized emission associated with the \hii\ regions - (a) and (b) 610 and 1280~MHz maps for the region associated with G346.056$-$0.021; (c) and (d) 610 and 1280~MHz emission for the region around G346.077$-$0.056. The contour levels are 3, 5, 10, 15, 20, 25, 30, 35 times of $\sigma$ with 
$\sigma$ is 0.3~mJy/beam and 0.2~mJy/beam at 610~MHz and 1280~MHz, respectively. Beam in each band is shown as filled ellipse. These maps are generated by setting
the `robustness parameter' to $-$5 and without any {\it uv} tapering.}
\label{radio_maps_rm5}
\end{figure*}

\begin{figure*}
\centering
\includegraphics[scale=0.25]{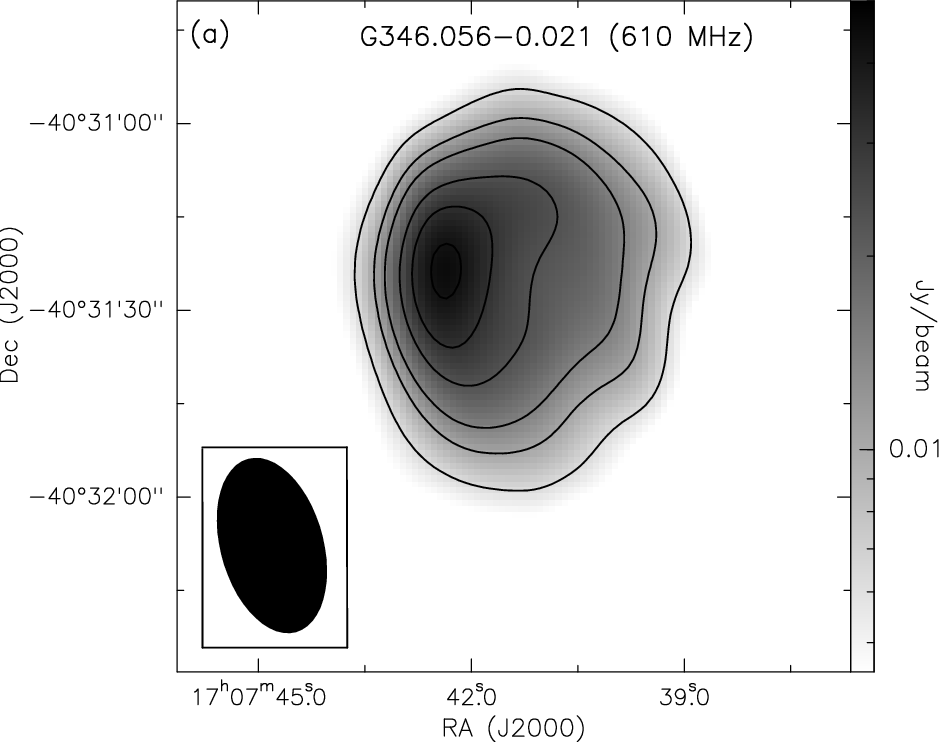}
\includegraphics[scale=0.25]{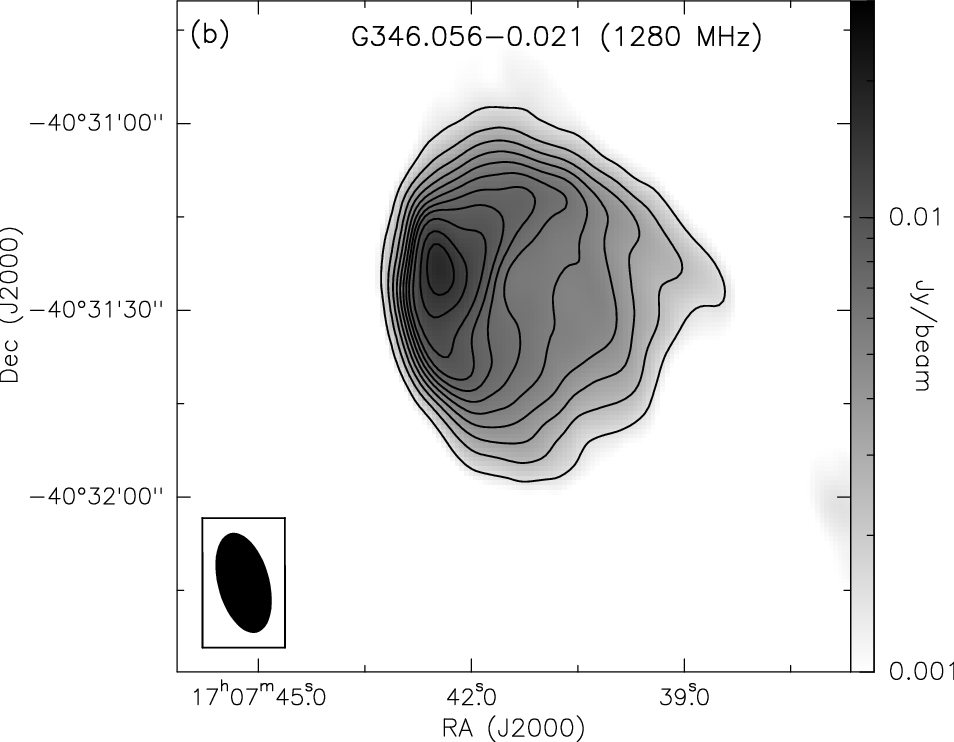}
\includegraphics[scale=0.25]{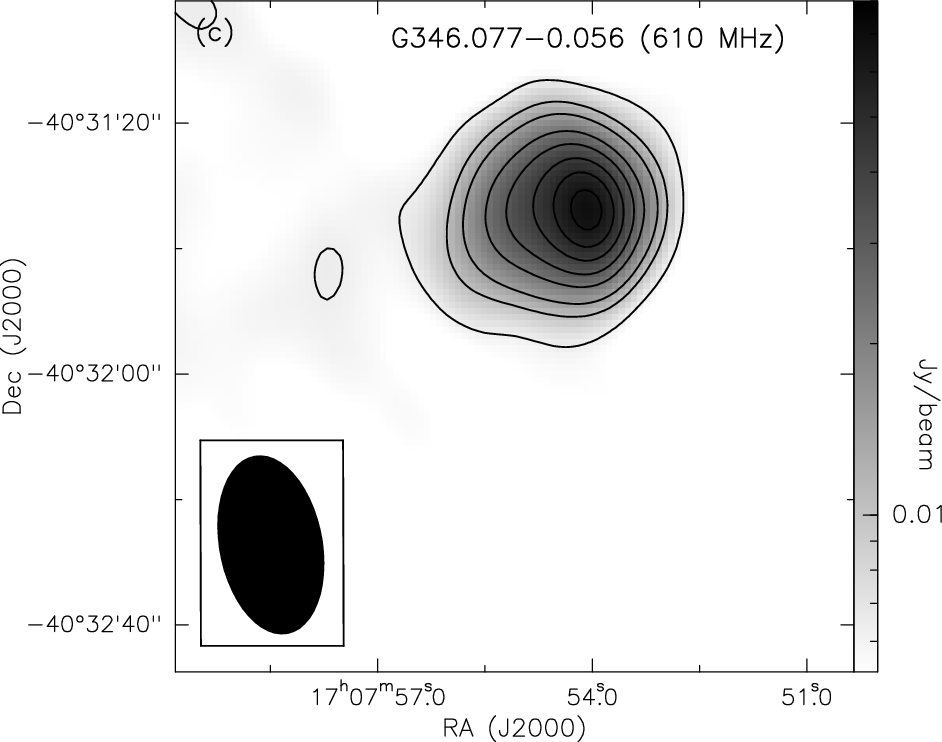}
\includegraphics[scale=0.25]{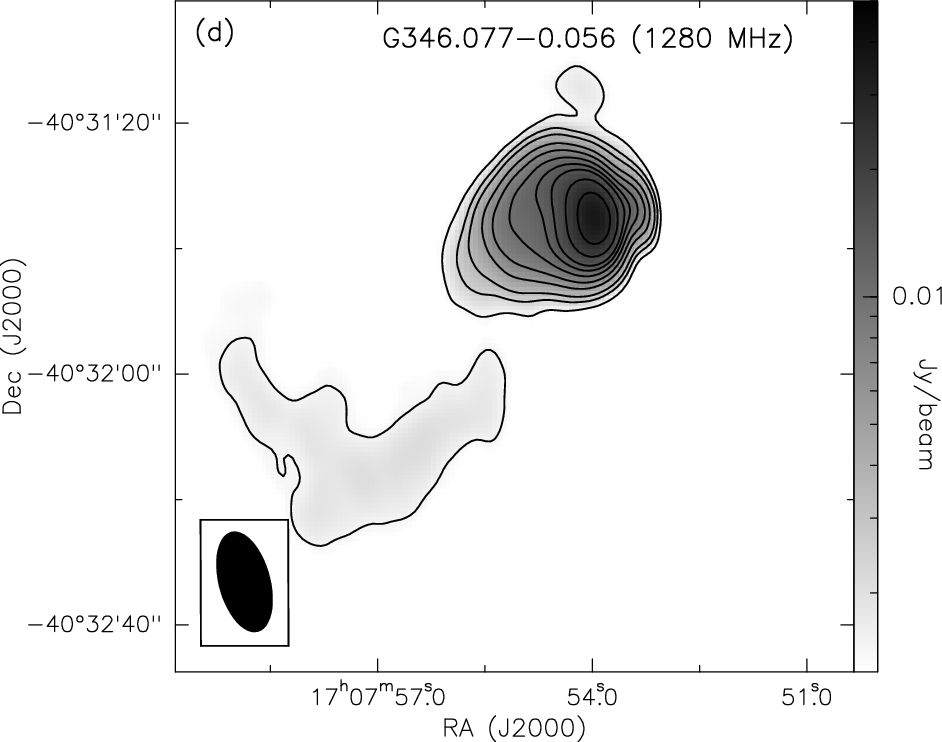}
\caption {Same as in Figure \ref{radio_maps_rm5}, but for maps generated with `robustness parameter' +1 and appropriate {\it uv} tapering to weigh down long baselines. The contour levels are 3, 5, 7, 11, 15, 20, 25, 30, 40, 55 times of $\sigma$ with 
$\sigma$ is 2.1~mJy/beam and 0.5~mJy/beam at 610~MHz and 1280~MHz, respectively. Beam in each band is shown as filled ellipse.}
\label{radio_maps_r1}
\end{figure*}

The radio emission associated with the two \hii\ regions mapped at 610 and 1280~MHz is
shown in Figure \ref{radio_maps_rm5}. These maps are generated by setting the
‘robustness’ parameter to $-$5 (on a scale where +5 represents pure natural weighting and $-$5 is for pure uniform weighting of the baselines) while running the task {\tt IMAGR} and considering the entire
{\it uv} coverage. However, for probing the larger spatial scales of the extended diffuse, ionized emission in these regions we also generate continuum maps by setting the ‘robustness’ 
parameter to +1 and weigh down the long baselines by using the task 
{\tt UVTAPER}. These lower resolution maps
are shown in Figure \ref{radio_maps_r1}.
The details of observation and the generated maps are listed in Table \ref{radio_tab}. The positional offsets of the peaks in these maps are
within $\sim$ 2\arcsec.
 
As seen from the figures, the ionized emission associated with the \hii\ region G346.056$-$0.021 displays a distinct cometary
morphology at both 610 and 1280~MHz with a steep intensity gradient towards the east. A faint, broad
and diffuse tail is seen towards the west-south-west. The observed morphology suggests that this \hii\ region is density bounded towards the south-west and ionization bounded towards the north-east.
The emission from G346.077$-$0.056 is also seen to be cometary 
in nature with the signature being more pronounced at 1280~MHz.  The higher resolution maps corroborate better with the above picture. The higher resolution 1280~MHz map displays an interesting morphology where a compact,
cometary structure is seen towards the west of G346.077$-$0.056. This could
possibly be another \hii\ region with its cometary head facing that of G346.077$-$0.056. However, we cannot rule out the other possibility of an externally ionized clump arising due to density inhomogeneities.
 An extended low-intensity (at 
3$\sigma$ level) detached component is also seen at 1280~MHz in Figure \ref{radio_maps_r1}. It is difficult to ascertain the physical association
of this feature with G346.077$-$0.056.
Using the Australia Telescope Compact Array, \citet{2007A&A...461...11U} have 
observed the two regions at 3.6~cm (8640~MHz) and 6.0~cm (4800~MHz) with a spatial resolution of $\sim$ 1 -- 2\arcsec~ and sensitivity $\sim$ 0.3 mJy.
Radio emission is detected at both frequencies for G346.077$-$0.056
with the peak positions in good agreement (within 2.5$\arcsec$) with the GMRT maps. They have classified this \hii\ region as one displaying a cometary morphogy
which is consonant with the GMRT maps. 
However, G346.056$-$0.021 is not listed under the RMS sources having associated
radio emission.
Table \ref{radio_res} allows
a comparison of GMRT results with that of \citet{2007A&A...461...11U} and \citet{1990ApJS...74..181Z}. The integrated flux densities derived from the GMRT 1280~MHz map is larger
compared to that from the 1400~MHz VLA observations obtained by \citet{1990ApJS...74..181Z}. This is understandable considering 
the fact that the integration time for their observation was only 120 seconds and
hence not expected to be sensitive to the extended, faint diffuse emission. 
Similarly, the high resolution ATCA maps would have also resolved out a good
fraction of the diffuse emission.

For deriving the physical parameters of the two \hii\ regions, we use the
lower resolution GMRT maps which samples most of the associated ionized emission.
For this we first convolve the 1280~MHz map to the resolution of the 610~MHz map.
From the peak flux densities at 610 and 1280~MHz, we estimate the
radio spectral index $\rm \alpha$ ($\rm F _\nu \propto \nu\ ^\alpha$) to be $-0.1\pm 0.06$ and $0.01\pm 0.004$ for G346.056$-$0.021 and G346.077$-$0.056, respectively.
Taking the integrated flux densities, we estimate spectral index
values of $-0.3\pm 0.06$ and $-0.4\pm 0.06$ for G346.056$-$0.021 and G346.077$-$0.056, respectively. Here, we have sampled the same region defined by the 3$\rm \sigma$ contour of the 610~MHz map. The spectral index values obtained from
the peak flux densities are consistent with optically thin free-free emission.
However, the values obtained from the integrated flux densities are indicative of
non-thermal emission \citep{{2016ApJS..227...25R},{1999ApJ...527..154K},{1993RMxAA..25...23R}}. Thus a scenario of co-existing free-free and non-thermal emission can be visualized for the \hii\ regions as has been addressed by 
several authors \citep{{2016A&A...587A.135R},{2016MNRAS.456.2425V},{2016AJ....152..146N},{2002ApJ...571..366M},{2017MNRAS.472.4750D}}. 
The above interpretation should be taken with caution because
GMRT is not a scaled array between the observed frequencies and the observed visibilities span different {\it uv} ranges. This implies that the generated
maps at 610 and 1280~MHz are sensitive to different spatial scales thus
rendering the estimated spectral indices uncertain. 

\begin{table}[h]
\caption{Details of the radio interferometric continuum observations and generated maps. Values in parenthesis are for the maps generated with `robustness parameter' $-$5 and no {\it uv} tapering.} 
\label{radio_tab}
\begin{tabular}{lll}
 
\hline\hline
Details & 610 MHz & 1280 MHz \\
\hline
Date of Obs. & 17 July 2011 & 20 July 2011 \\
Flux Calibrators & 3C286,3C48 & 3C286,3C48\\
Phase Calibrators & 1626$-$298 & 1626$-$298\\
On-source Integration time & $\sim$~4~hr & $\sim$~4~hr \\

Synth. beam & 14.4\arcsec$\times$8.5\arcsec & 8.8\arcsec$\times$4.4\arcsec \\
            &  (6.4\arcsec$\times$3.1\arcsec) & (3.6\arcsec$\times$1.5\arcsec)  \\
Position angle. (deg) & 10.6 & 15.0 \\            
                      &  (6.6) &  (3.3) \\
                      
{\it rms} noise (mJy/beam) & 2.1 & 0.5 \\
                           &  (0.3) & (0.2) \\
\hline
\end{tabular}
\end{table}

\begin{table*}
\centering
\tiny
\caption{GMRT, ATCA \citep {2007A&A...461...11U} and 1.4~GHz \citep{1990ApJS...74..181Z} results. The peak coordinates (from 1280~MHz map$^\star$), peak and integrated flux densities of the two \hii\ regions are listed. The integrated flux density in GMRT maps are calculated by integrating above 3$\rm \sigma$ level.
 Values for the convolved 1280~MHz map are listed in the second line. For the 
integrated flux densities, the area probed is kept same as in 610~MHz. Values in parenthesis are from the radio maps, generated by setting the `robustness parameter' to $-$5 and no {\it uv} tapering.}
\label{radio_res} 
\begin{tabular}{cccccccccccc}
\\ \hline
 \multicolumn{2}{c}{\underline{~~~~~~~~Peak Coordinates~~~~~~~~}} & \multicolumn{5}{c}{\underline{~~~~~~~~~~~~~~~~~~~~~~~~~~~~~~~~Peak flux (mJy/beam)~~~~~~~~~~~~~~~~~~~~~~~~~~~~~~~~}} & \multicolumn{5}{c}{\underline{~~~~~~~~~~~~~~~~~~~~~~~~~~~~~~~~~~~~Integrated flux (mJy)~~~~~~~~~~~~~~~~~~~~~~~~~~~~~~~~~~~}} \\

RA (J2000)  & DEC (J2000) & 610~MHz & 1280~MHz & 1400~MHz & 4.8~GHz & 8.6~GHz & 610~MHz$^\ddagger$ & 1280~MHz$^\ddagger$ & 1400~MHz & 4.8~GHz & 8.6~GHz \\
\hline 
\multicolumn{12}{c}{G346.056$-$0.021} \\
\hline
17:07:42.50 & $-$40:31:23.00 & 44.7 & 17.3 & 22 & -- & -- & 339$\pm$34 & 271$\pm$27 & 101 & -- & -- \\ 
 &  &  &  40.4 &  &  &  &  & 268$\pm$27 &  &  &  \\ 
&  &  (11.37) &  (4.02) &  &  &  &  (245$\pm$24.5) & (178$\pm$18) &  &  &  \\ 
\hline
\multicolumn{12}{c}{G346.077$-$0.056} \\
\hline
17:07:54.00 & $-$40:31:35.40 & 67.8 & 35.7  & 36 & 8.3 & 3.2 & 225$\pm$22 & 173$\pm$17  & 62 & 46.2 & 15.7 \\
&  &  &  68.6 &  &  &  &  & 174$\pm$17 &  &  &  \\ 
 &  & (23.7) & (8.47) &  & &  &  (197$\pm$19) & (159$\pm$16) &  &  &  \\ 
\hline
\end{tabular} 
\\$^\star$ The peak positions of the GMRT maps are consistent with
the 1400~MHz (VLA) map (within 5.5\arcsec) and the 4.8 and 8.6~GHz (ATCA) maps (within 2.5\arcsec).
\\ $^\ddagger$ Error in integrated flux has been calculated following the equation from 
\citet{2013ApJ...766..114S},
$\rm (2\sigma(\theta_{source}/\theta_{beam})^{1/2})^2 + (2\sigma_{flux--scale})^2 ]^{1/2}$, where $\sigma$ 
is the rms noise level of the map, $\rm \theta_{source}$ and $\rm \theta_{beam}$ are the size of the source 
and the beam, respectively, and $\rm \sigma_{flux--scale}$ is the error in the flux scale, which takes into 
account the uncertainty on the calibration applied to the integrated flux of the source. For GMRT maps 
uncertainty in the flux calibration is 5\% \citep{2007MNRAS.374.1085L}
\end{table*}

Following the method discussed in  \citet{{2006ApJ...653.1226Q},{2015ApJ...810...42A},{2016ApJ...824..125L}}, we derive the electron temperature,  $T_{ e}$, towards both these regions. This formulation assumes local thermodynamic equilibrium for the RRL lines, and is
given by the following expression
\begin{equation}
\label{electron_temp}
\begin{split}
T_{e}[K] =  \left \{ 7103.3 \left (\frac{\nu}{\rm GHz}\right )^{1.1}  \left (\frac{T_C}{T_L(\rm H^+)}\right ) \left( \frac{\Delta V(\rm H^+)}{\rm km\ s^{-1}} \right )^{-1} \right. \\ 
& \hspace{-2.5cm} \left. \rm  \times \left( 1+ \frac{\rm n(^4He^+)}{n(H^+)} \right )^{-1} \right \} ^{0.87}
\end{split} 
\end{equation}
where, $\nu$ is the observing frequency for the RRL lines, $T_C$ is the peak continuum antenna temperature, $T_L$ is the peak antenna temperature for hydrogen RRL line, $\Delta V(\rm H^+)$ is the FWHM line width for the RRL line and $\rm n(^4He^+)/n(H^+)= y^+$ is the helium ionic abundance ratio. The hydrogen RRL line parameters for both the \hii\ regions are taken from \citet{2011ApJS..194...32A}. 
 The helium ionic abundance ratio has been derived from the hydrogen and helium RRL line properties and using the following equation \citep{{2006ApJ...653.1226Q},{2013ApJ...764...34W}}   
\begin{equation}
\label{ionic_abundance}
{\rm y^+}= \frac{T_L(\rm ^4He^+) \Delta V(\rm ^4He^+)}{T_L(\rm H^+)\Delta V(\rm H^+)}
\end{equation}
where, $T_L$ is the peak line intensity and $\Delta V$ is the FWHM line width. We have derived the value of $\rm y^+$ to be 0.11 for the \hii\ region G346.056$-$0.021 using the 
hydrogen and helium RRL line parameters from \citet{2013ApJ...764...34W}. For G346.077$-$0.056, no helium RRL observation is available, hence we use an average value of $\rm y^+$= 0.07 estimated from a sample of \hii\ regions \citep{2013ApJ...764...34W}. 
From the above expressions and observed parameters, we estimate the electron temperature to be 5500~K and 8900~K for the \hii\ regions G346.056$-$0.021 and G346.077$-$0.056, respectively. These values fall within the range
of $\sim5000$ to $\sim10000$~K seen for Galactic \hii\ regions \citep{2006ApJ...653.1226Q}.

In order to derive the other physical properties of the ionized emission associated with the 
two \hii\ regions, we adopted the expressions from 
\citet{2016A&A...588A.143S}. Lucid explanations coupled with rigorous 
derivations of physical properties of \hii\ regions can be found in the
original papers of  \citet{{1967ApJ...147..471M},{1968ApJ...153..761R},{1969ApJ...156..269S},{1973AJ.....78..929P}}. The \hii\ regions are considered as Str\"{o}mgren's spheres which
are fully ionized spherical regions of uniform electron density. Assuming the radio emission at 1280~MHz to be optically thin and emanating
from a homogeneous, isothermal medium, the electron density, $n_{\rm e}$, 
the emission measure, EM, and the number of Lyman-continuum photons per second, $N_{\rm Ly}$, 
are estimated using the following equations \citep{2016A&A...588A.143S} 

\begin{equation}
\left( \frac{\rm EM}{\rm pc\ cm^{-6}}\right) = 3.217 \times 10^7  \left(\frac{\rm F_\nu}{\rm Jy}\right) \left( \frac{T_e}{\rm K}\right)^{0.35} \left( \frac{\nu}{\rm GHz}\right)^{0.1} \left( \frac{\theta_{\rm source}}{\rm arcsec}\right)^{-2}
\end{equation}
\begin{equation}
\begin{split}
\left( \frac{n_e}{\rm cm^{-3}}\right)  = 2.576 \times 10^6 \left(\frac{\rm F_\nu}{\rm Jy}\right)^{0.5} \left( \frac{T_e}{\rm K}\right)^{0.175} \left( \frac{\nu}{\rm GHz}\right)^{0.05} \\
 & \hspace{-2.5cm}  \times \left( \frac{\theta_{\rm source}}{\rm arcsec}\right)^{-1.5} \left( \frac{\rm D}{\rm pc}\right) ^{-0.5}
\end{split}
\end{equation}
\begin{equation}
\left( \frac{N_{Ly}}{\rm Sec^{-1}}\right) = 4.771 \times 10^{42} \left(\frac{\rm F_\nu}{\rm Jy}\right) \left( \frac{T_e}{\rm K}\right)^{-0.45} \left( \frac{\nu}{\rm GHz}\right) \left( \frac{\rm D}{\rm pc}\right) ^{2}
\end{equation}
where, $F_{\nu}$ is the integrated flux density of ionized region, $T_e$ is the electron temperature, $\nu$ is the frequency, $\theta_{\rm source}$ is the angular diameter of the \hii\ region, and D is the distance to these regions. 
The angular extents of the ionized emission associated with G346.056$-$0.021 and G346.077$-$0.056 are 
estimated from the 1280~MHz map to be $61\arcsec \times 49\arcsec$ and $34\arcsec \times 32\arcsec$, respectively. These
values are the $\rm FWHM_x \times FWHM_y$ obtained using the 2D {\it Clumpfind} algorithm \citep{1994ApJ...428..693W} for a $3\sigma$ threshold level. 
Applying beam corrections, we derive the deconvolved sizes to be $33\arcsec \times 38\arcsec$ and $17\arcsec \times 19\arcsec$ for G346.056$-$0.021 and G346.077$-$0.056, respectively. For
$\theta_{\rm source}$, we have taken the geometric mean of the deconvolved $\rm FWHM$.
The physical parameters thus derived are listed in Table \ref{radio_param}. 
 
\begin{table*}
\centering
\caption{Derived physical parameters of \hii\ regions.} 
\label{radio_param}
\begin{tabular}{cccccccc}
\hline 
Source & $T_e$  & $\theta_{\rm source}$ & $n_e$ $\rm $ & EM $\rm $ & log $ N_{Ly}$ & Spec. Type & $\rm t_{dyn} $\\
& (K)& (arcsec) & ($\rm cm^{-3}$) & $\rm (cm^{-6}\ pc)$ & & & (Myr) \\ 
\hline
G346.056$-$0.021 & 5500  & 35 & 4.2$\times 10^2$ & $2.5 \times 10^5$ & 48.50 & O7.5V - O7V & 0.5 \\
G346.077$-$0.056 & 8900  & 18 & 1.0$\times 10^3$ & $7.1 \times 10^5$ & 48.21 & O8.5V - O8V & 0.2 \\
\hline
\end{tabular}
\end{table*} 

If single ZAMS stars are responsible for the ionization of the \hii\ regions, then 
using Table I of \citet{2005A&A...436.1049M}, the
estimated Lyman continuum flux translates to spectral types O7.5V -- O7V and 
O8.5V -- O8V for
G346.056$-$0.021 and G346.077$-$0.056, respectively. Similar spectral types are obtained if we use the
results of \citet{2011MNRAS.416..972D} and \citet{2011ApJ...730L..33M}.
 If we consider the compact, cometary \hii\ region seen in the higher
resolution 1280~MHz map to be internally ionized, then the integrated flux density
implies a massive star of spectral type B0.5 -- B0. The spectral type 
estimate for the ionizing star of G346.077$-$0.056 remains same if we subtract
out the flux density of this component.
Taking the bolometric luminosities from the RMS survey paper by \citet{2013ApJS..208...11L} and comparing the same with the tables of \citet{2011ApJ...730L..33M}, we obtain
consistent spectral type estimates of O8 -- O7.5 for both the \hii\ regions.
As mentioned earlier, this estimate is with the assumption of optically thin
emission and hence serves as a lower limit as the emission could be optically thick at 1280~MHz. Several studies have shown that dust absorption of Lyman
continuum photons can be very high \citep{{2001ApJ...555..613I},{2004ApJ...608..282A},{2011A&A...525A.132P}}. 
With limited knowledge of the dust properties, we have not accounted for the dust absorption 
in the above estimates. The Lyman continuum fluxes suggest massive stars of 
masses $\rm \sim 25$ and $\rm \sim 20 M_{\odot}$ responsible for the ionized emission of G346.056$-$0.021 and G346.077$-$0.056, respectively \citep{2011MNRAS.416..972D}. 

We use a simple model discussed in \citet{{1978ppim.book.....S},{1980pim..book.....D}} to 
estimate the dynamical ages of the two \hii\ regions. 
If an \hii\ region evolves in a homogeneous medium then its dynamical age can be 
estimated from the following expressions
\begin{equation}
\rm R_{st} = \left[\frac{3\it N_{Ly}}{4\ \pi\ n_{H,0}^{2}\ \alpha_{B}} \right]^{1/3}
\label{stromgren}
\end{equation}
\begin{equation}
\rm t_{dyn} = \frac{4}{7}\ \frac{R_{st}}{C_{Hii}} \left[ \left(\frac{R_{if}}{R_{st}}\right)^{7/4}\ -1\right]
\end{equation}
where, $\rm R_{st}$ is the Str\"omgren radius, $ N_{Ly}$ is the Lyman continuum photons coming 
from the ionizing source, $\rm n_{H,0}$ is the particle density of the neutral gas, 
$\rm \alpha_{B}$ is the coefficient of radiative recombination and is taken to
be 2.6 $\times 10^{-13}$ $\rm (10^{4}\ K/T)^{0.7}$ $\rm cm^{3}\ 
sec^{-1}$ from \citet{1997ApJ...489..284K}. In the second expression, $\rm t_{dyn}$ is the dynamical age, $\rm C_{Hii}$ is the 
isothermal sound speed of ionized gas (assumed to be 10~$\rm km\ s^{-1}$), and $\rm R_{if}$ is the 
radius of the \hii\ region. For $\rm R_{if}$, we use the deconvolved sizes of
17$\arcsec$~(0.9~pc) and 9$\arcsec$~(0.5~pc) for G346.056$-$0.021 and G346.077$-$0.056, respectively. $\rm n_{H,0}$ 
is estimated from the column density map (refer to Section \ref{dust}) and is found to be
$\rm 6.2 \times 10^4 cm^{-3}$ and $\rm 4.8 \times 10^4 cm^{-3}$ for G346.056$-$0.021 and G346.077$-$0.056, respectively. For G346.077$-$0.056, the estimated values of $\rm n_{H,0}$ is higher by a 
factor of $\sim$ 6 compared with
that obtained by \citet{2015MNRAS.446.2566Y}.
Using these values in the above equations, we estimate the dynamical ages for G346.056$-$0.021 and
G346.077$-$0.056 to be 0.5 and 0.2~Myr, respectively. The derived age should however be treated with caution
since the assumption of expansion in a homogeneous medium is not a realistic one.

\subsection{Associated stellar population}
 
\begin{figure*}[t]
\centering
\includegraphics[scale=0.25]{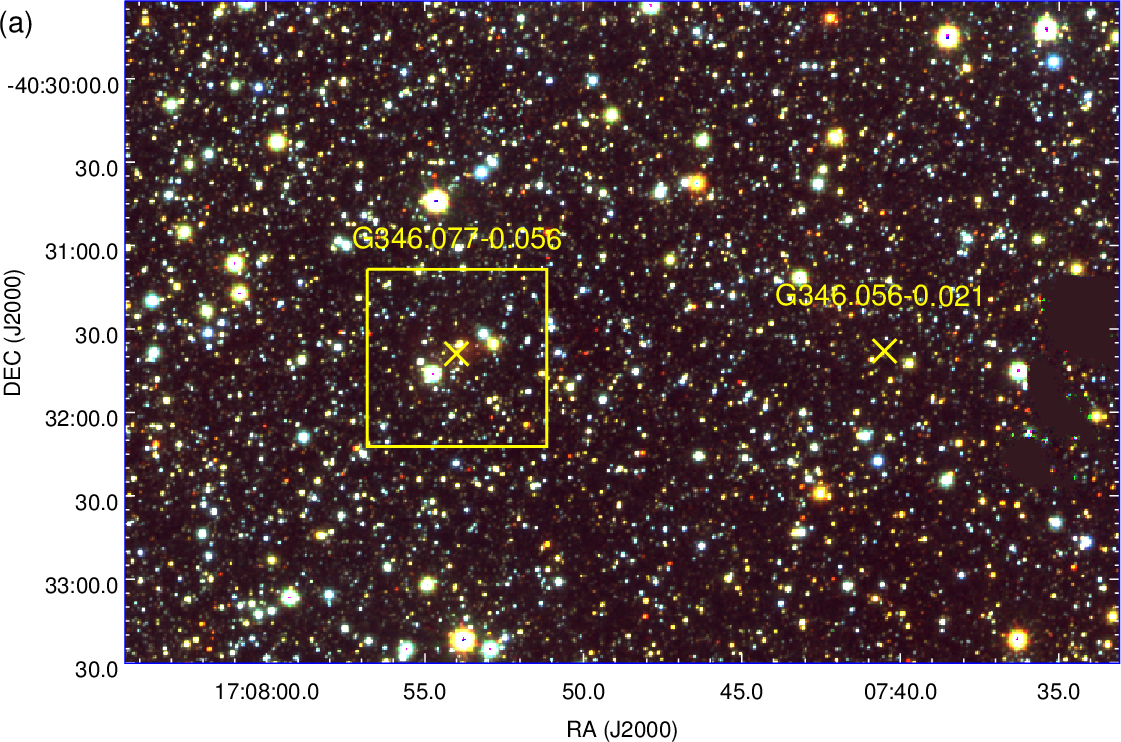}
\includegraphics[scale=0.25]{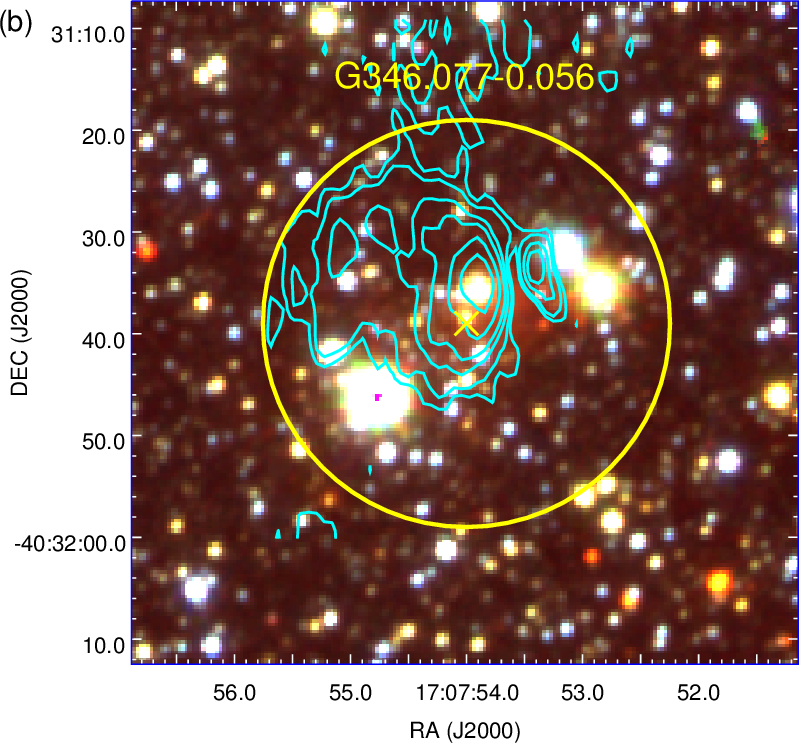} 
\caption {(a) NIR colour composite image of the region associated with the \hii\ regions using VVV JHK band images (red -- K; green -- H; blue -- J). $\times$' marks show the location of the \hii\ regions. (b) Zoomed in view of indicated region related to G346.077$-$0.056.
The yellow circle denotes
the size of the cluster as estimated by \citet{2011A&A...532A.131B}. 
 The cyan contours show the higher resolution 1280~MHz radio emission with the levels same as those plotted in Figure \ref{radio_maps_rm5}.}
\label{JHK_comp_VVV}%
\end{figure*}

\begin{figure*}[t]
\centering
\includegraphics[scale=0.3]{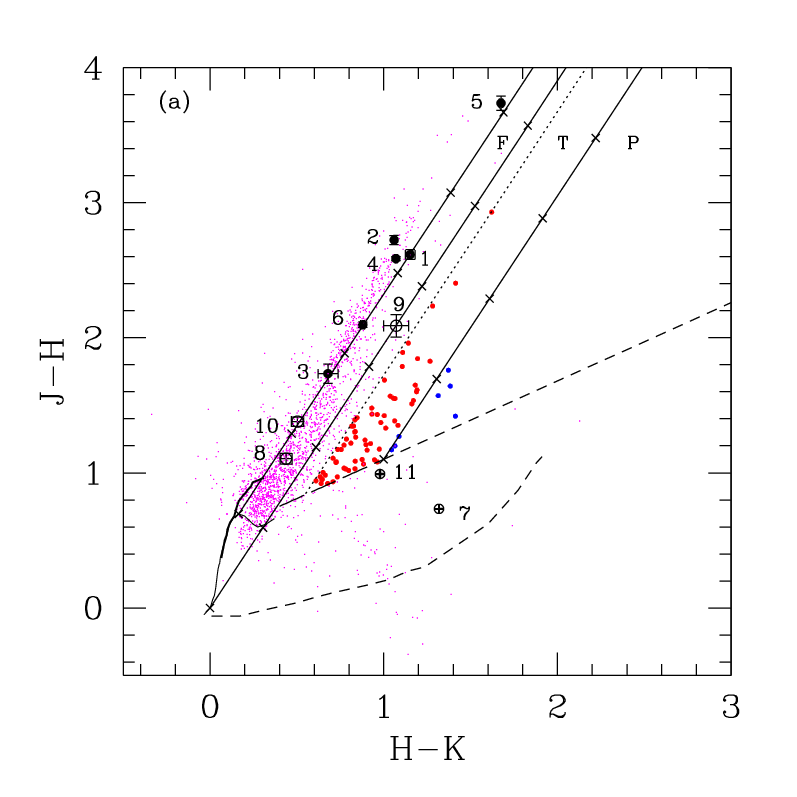}
\includegraphics[scale=0.3]{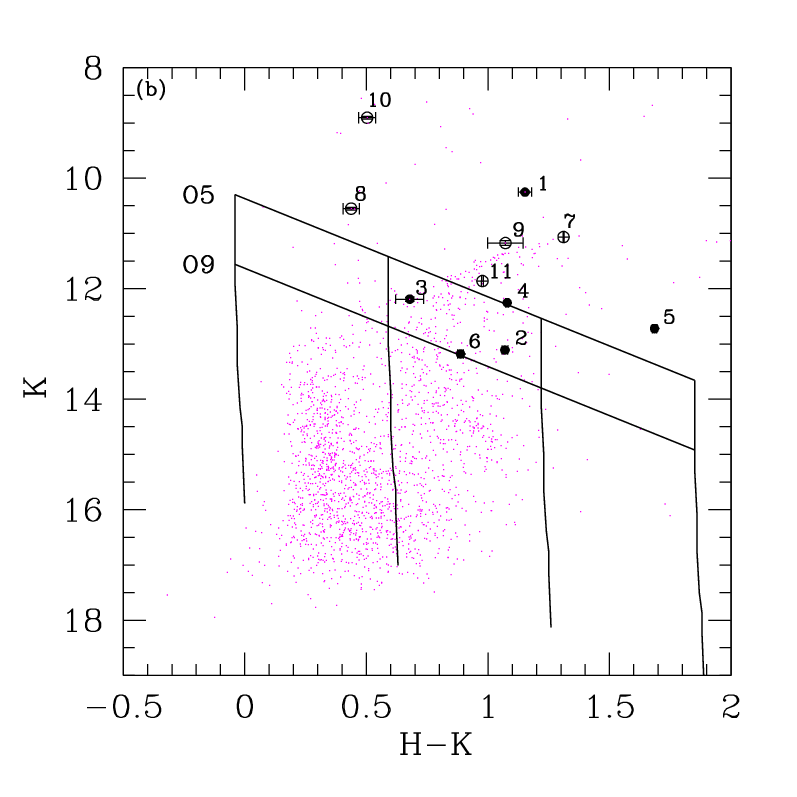}
\caption {(a) (J $-$ H) vs (H $-$ K) CCP for the region associated with the \hii\
regions. The loci of main sequence (thin line) and giants (thick line) are taken from \citet{1988PASP..100.1134B}. The classical T Tauri locus (long dashed line) is
adopted from \citet{1997AJ....114..288M} and that for the Herbig AeBe stars (short dashed line)
is from \citet{1992ApJ...393..278L}. The parallel lines are the reddening vectors
where cross marks indicate intervals of 5 mag of visual extinction.
The interstellar reddening law assumed is from \citet{1985ApJ...288..618R}. The colours and curves
in the CCP are all converted into \citet{1988PASP..100.1134B} system. 
The regions `F', `T' and `P' are discussed in the text. The dotted line parallel
to the reddening vector accounting for an offset of three times the photometric error in the bands. On the CCP, the Class I sources (blue), Class II sources (red) sources  are shown as filled circles.
The candidate ionizing stars are shown as filled black circle (for G346.056$-$0.021) and open circles (for G346.077$-$0.056) on both CCP and CMP. The individual error bars
on the colours and magnitude are also plotted. 
(b) K vs (H $-$ K) CMP for the region associated with the \hii\ regions. The nearly vertical solid lines represent the ZAMS loci with 0, 10, 20 and 30 magnitudes of visual extinction corrected for the distance. The slanting lines show the reddening
vectors for spectral types O9 and O5. The magnitudes and the ZAMS loci are all plotted in the Bessell \& Brett (1988) system.}
\label{CC_CM_plot}%
\end{figure*}

Figure \ref{JHK_comp_VVV} shows the NIR view of the region associated with the
two \hii\ regions. The region is seen to be densely populated. 
The zoom of the region associated with G346.077$-$0.056 shows the presence of
faint K-band nebulosity harbouring the IR cluster VVVCL094 \citep{2011A&A...532A.131B}. These authors have estimated the cluster radius to be 20\arcsec. After
a statistical decontamination of field star population, they propose 20
probable members for this cluster. 
As part of another statistical study of clusters in the inner Galaxy, \citet{2013A&A...560A..76M} have classified VVVCL094 as an embedded cluster whose estimated centre
coincides with the ATLASGAL peak. Given the correlation of the cluster with the
probed ionized region, it is likely that the most massive members of it are
responsible for the detected \hii\ region. 

Using the NIR data from the VVV and 2MASS surveys, we attempt to identify
candidate ionizing star(s) responsible for the \hii\ regions and study the
distribution of the associated YSOs. We probe the region shown in Figures \ref{region_8_24_850} and \ref{JHK_comp_VVV}. 
Following the criteria outlined in \citet{2013IJAA....3..161F},
we retain sources satisfying {\tt mergedClass} values of $-$1 or $-$2, {\tt pstar} $\ge$ 
0.999656 and {\tt priOrSec} set to 0 or {\tt frameSetID}. This
ensures that the retrieved sample contains stellar objects with good quality
photometry. Given the VVV saturation limit of 11 mag in the JHK bands, we 
include sources with good quality photometry (``read-flag" = 2) from the 
2MASSPSC brighter than 11 mag. Further, in our sample we have
included a few sources fainter than 11 mag from the 2MASSPSC
which do not have good quality VVV photometry in all bands. Adopting the set of
criteria described above we generate a merged catalog of 2013 sources with
1927 and 86 sources from the VVV catalog and the 2MASSPSC, respectively. 

\begin{table*}
\centering
\caption{\small Details of the candidate ionizing star(s) for the \hii\ regions.}
\label{exciting_table}
\begin{tabular}{lccccc}
\\ \hline 
\# & RA (J2000) & DEC (J2000) & J & H & K / K$_s$ \\
 &(hh:mm:ss.ss)&(dd:mm:ss.ss)&(mag)&(mag)&(mag) \\
\hline 
1$^*$ & 17:07:39.81 & $-$40:31:41.43 & 13.91 & 11.37 & 10.21 \\
2 & 17:07:40.16 & $-$40:31:51.95 & 16.57 & 14.22 & 13.11 \\
3$^*$ & 17:07:40.46 & $-$40:31:11.82 & 14.51 & 12.84 & 12.15  \\
4 & 17:07:40.99 & $-$40:31:39.97 & 15.61 & 13.37 & 12.25  \\
5 & 17:07:41.65 & $-$40:31:24.00 & 17.71 & 14.46 & 12.73 \\
6 & 17:07:43.02 & $-$40:31:33.92 & 15.90 & 14.09 & 13.17 \\
7$^*$ & 17:07:52.93 & $-$40:31:34.40 & 12.96 & 12.36 & 11.05 \\
8$^*$ & 17:07:53.25 & $-$40:31:30.87 & 12.02 & 10.97 & 10.51 \\
9$^*$ & 17:07:53.98 & $-$40:31:34.85 & 14.23 & 12.21 & 11.13 \\
10 & 17:07:54.85 & $-$40:31:45.17 & 10.71 & 9.39 & 8.87 \\
11 & 17:07:55.24 & $-$40:31:40.80 & 13.67 & 12.84 & 11.84 \\
\hline 
\end{tabular}
\\ $^*$ Photometric magnitude are from 2MASSPSC.
\end{table*}

\begin{figure*}[t]
\centering
\includegraphics[scale=0.4]{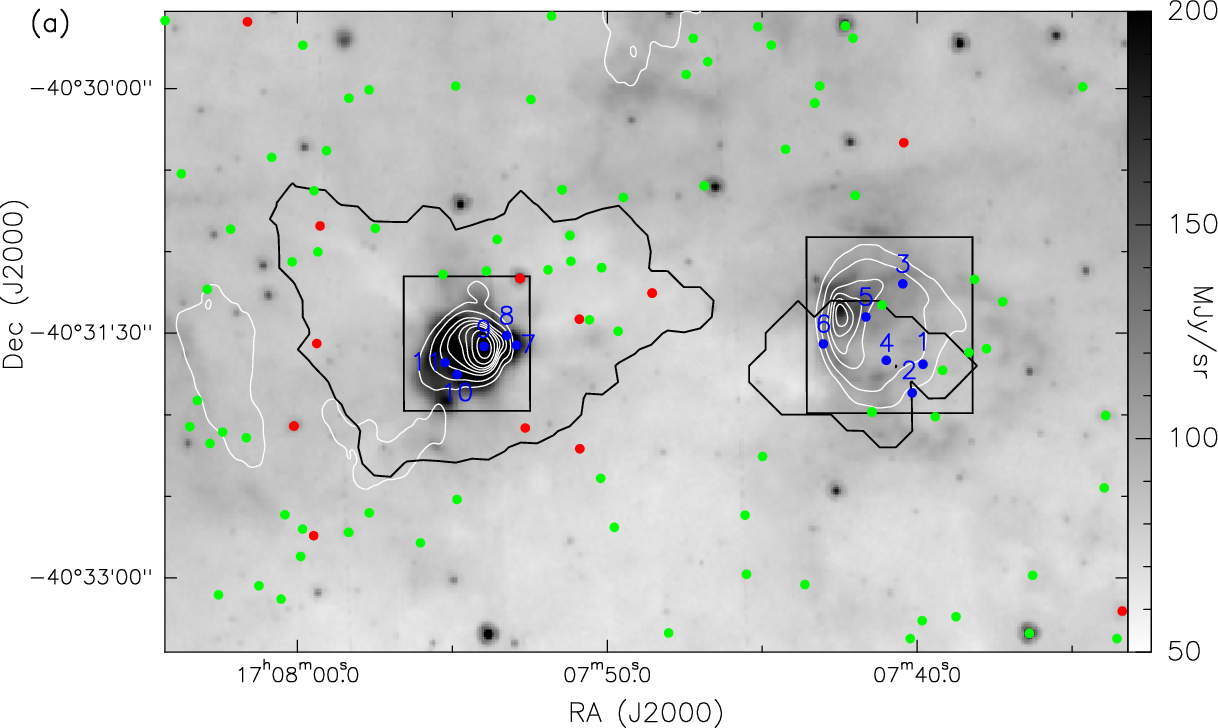}
\hspace{-3cm}
\includegraphics[scale=0.31]{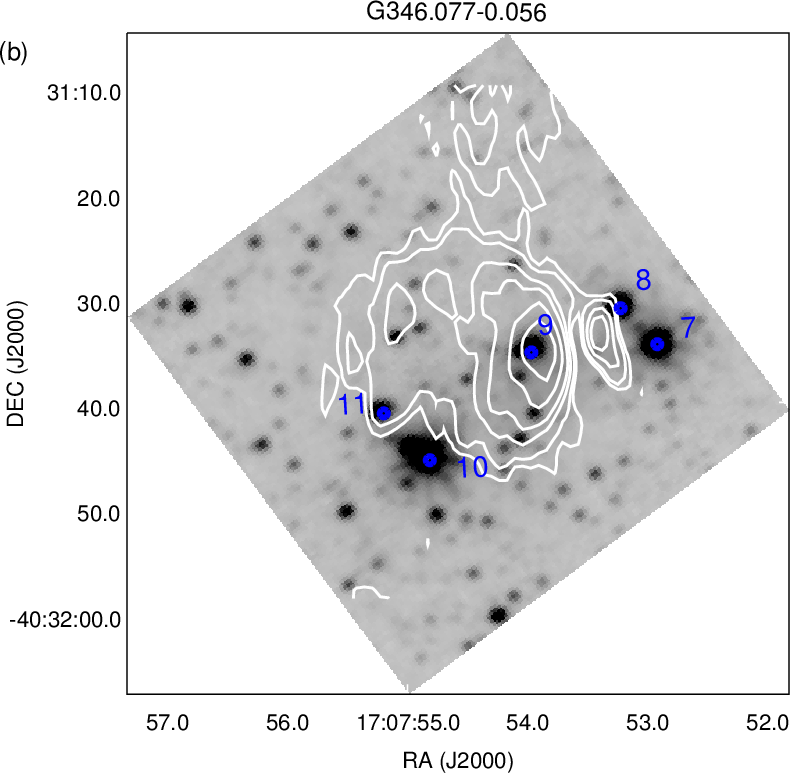}
\includegraphics[scale=0.31]{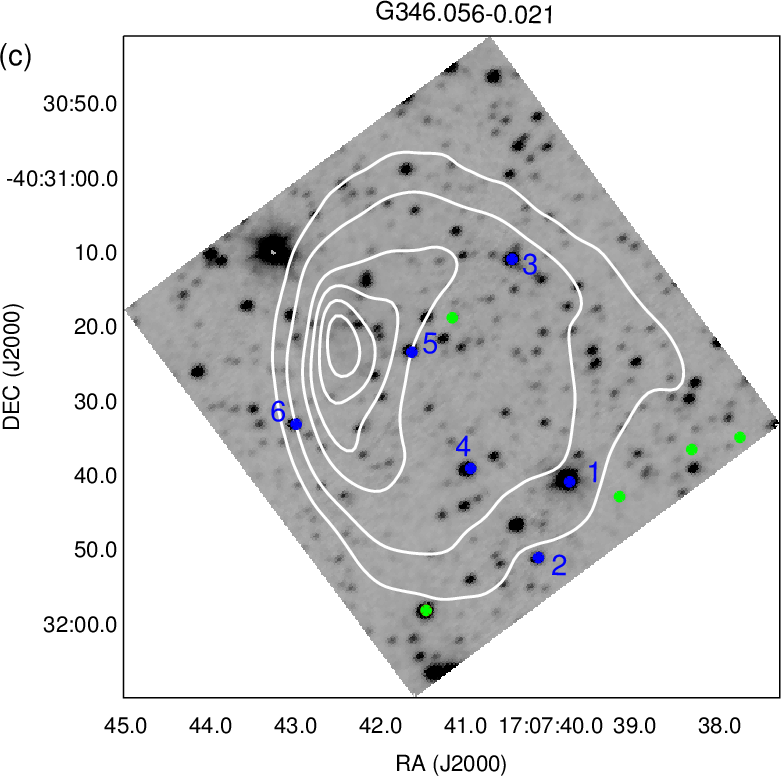}
\caption {(a) Distribution of candidate ionizing stars and YSOs on the {\it Spitzer} 8.0~$\rm \mu m$ image. Identified YSOs have the following colour coding - Class I (red), Class II (green). The ionizing stars are shown in blue. The detected clump apertures (see
Section \ref{dust} are displayed in black. 1280 MHz (low resolution) radio contours are overlaid with levels of 3, 9, 13, 25, 30, 40, 50 times of $\sigma$. (b) and (c) are zooms of regions 
(marked as black rectangles in (a))
associated with G346.077$-$0.056 and G346.056$-$0.021, respectively on the K-band VVV image. In (b), we have overlaid the high resolution 1280~MHz radio contours shown in Figure \ref{radio_maps_rm5} to reveal the finer structures.} 
\label{yso_excitng_dist}%
\end{figure*}

Figure \ref{CC_CM_plot} shows the (J $-$ H) vs (H $-$ K) colour-colour plot (CCP) and K vs (H $-$ K) colour-magnitude plot (CMP) for the sources in our merged catalog. 
In our search for the candidate ionizing stars, we take into account two points - (i) the ionizing stars are likely to be located within the radio contours and 
(ii) the Lyman continuum photon flux estimates from the GMRT radio maps (see Section \ref{ionized}) sets lower limits on the spectral types of O7.5V -- O7V and O8.5V -- O8V for G346.056$-$0.021 and G346.077$-$0.056, respectively. 
Hence, on the CCP and CMP we highlight the sources falling within the
$3\sigma$ contours of the radio emission for G346.056$-$0.021 and G346.077$-$0.056, respectively and also lying above the reddening vector for
spectral type O9. The identified sources are labeled 
\# 1 to \# 6 (black filled circles) for G346.056$-$0.021 and \# 7 to \# 11 
(black open circles) for G346.077$-$0.056. Table \ref{exciting_table} lists
the position and photometric magnitudes of these sources and Figure \ref{yso_excitng_dist}(a) shows
the location of these sources on the MIR 8~$\rm \mu m$ IRAC image with overlaid radio contours. The ATLASGAL clumps, discussed in the next section,
are also shown in this figure. Figures \ref{yso_excitng_dist}(b) and (c) show the
zoomed in region for G346.077$-$0.056 and G346.056$-$0.021, respectively, on the
K-band VVV image.

As seen from the CCP, except sources \# 7 and \# 11 the rest of the likely candidates earlier than O9 fall in the region occupied by main-sequence or Class III sources. The location of sources \# 7 and \# 11 indicate that they
are mostly reddened Herbig AeBe stars. 
Further, given the distance of 10.9~kpc to the \hii\ regions and assuming
a foreground extinction of
$A_V \sim 1$ per kpc, the CMP suggests that sources \# 8 and \# 10 are 
unlikely to be associated with the \hii\ regions. This makes the source \# 9, which coincides
with the radio peak (within 1$\arcsec$) a promising candidate ionizing star responsible for G346.077$-$0.056.  In case of G346.056$-$0.021, all the early type stars within the
ionized region lie close to the left-most reddening vector. Within the photometric errors and the uncertainty involved in defining the reddening vector itself,
sources \# 1, \# 3, \# 5, and \# 6 maybe considered as Class III sources. However,
one cannot rule out the possibility of these being highly reddened giants. Looking
at their distribution within the ionized region, source \# 5 has a better chance
of qualifying as the ionizing source. This corroborates well with the fact that
 G346.056$-$0.021 is a young, compact \hii\ region and thus one expects the ionizing
 star to be close to the radio peak. Another likely candidate would be the Class II
source (located at $\alpha_{2000}$=17:07:41.13, $\delta_{2000}$=$-$40:31:19.67 with J = 17.16, H = 16.14, and K = 15.22) located close to source \# 5. This can
also be considered as a massive, embedded exciting source. In their study of the
IR dust bubble S51, \citet{2012A&A...544A..11Z} 
identified a Class II massive (O-type) source as the candidate ionizing star. 
Detailed spectroscopy and spectral energy distribution modeling is required to
confirm the identified candidates.
    
\begin{figure*}[t]
\centering
\includegraphics[scale=0.3]{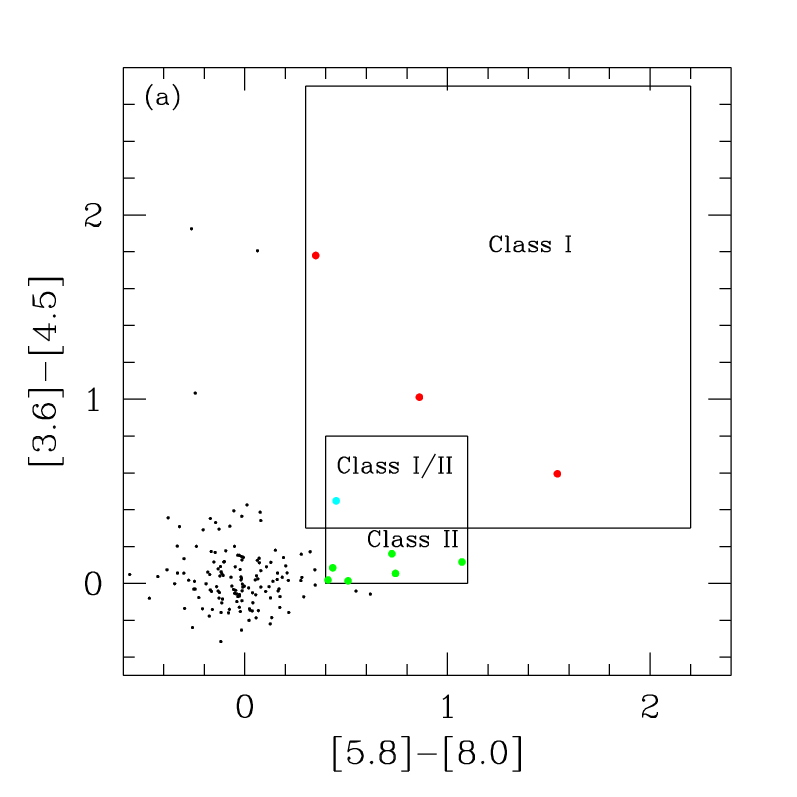}
\includegraphics[scale=0.3]{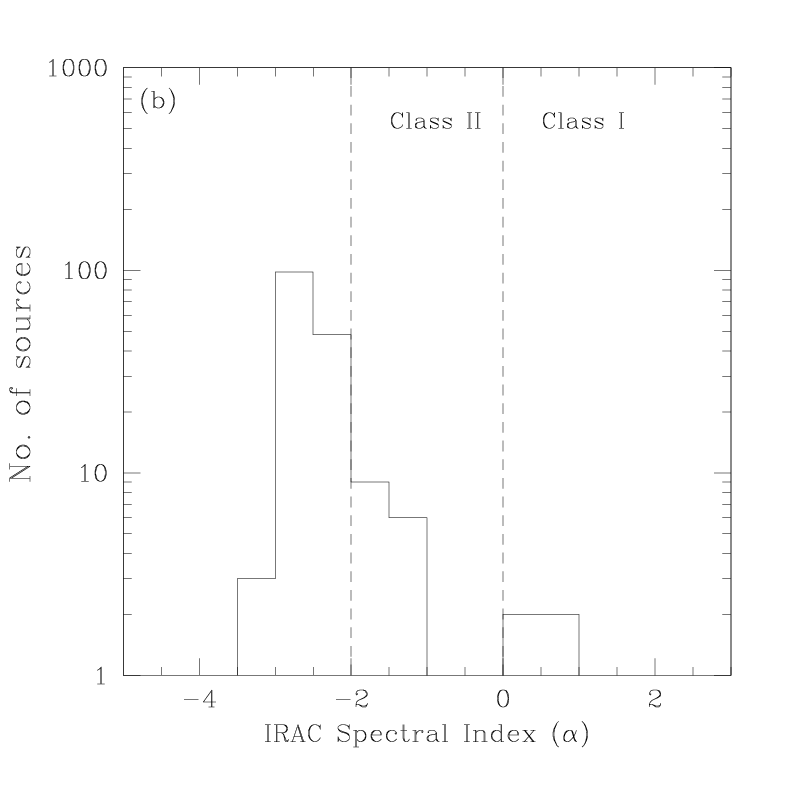}
\caption{(a) IRAC colour-colour plot for the sources in the \hii\ regions. The boxes demarcate the location of Class I (larger box) and Class II (smaller box) \citep{2007A&A...463..175V}. Sources falling in the overlapping area are designated as Class I/II. The identified YSOs have the following colour coding - Class I (red), Class II (green) and Class I/II (cyan). (b) The histogram showing the number of sources within specified 
spectral index bins. The regions demarcated on the plot are adopted from \citet{2008ApJ...682..445C} for classification of YSOs.}
\label{irac_yso}%
\end{figure*}

To understand the star formation activity towards the \hii\ regions,
we examine the spatial distribution of YSOs using the {\it Spitzer} GLIMPSE, VVV and 
2MASS point source photometry. The various schemes adopted for YSO identification are outlined below.
\begin{enumerate}
\item Here, we have used the classification criteria discussed in \citet{2004ApJS..154..363A}. 
In this method, the Class I (sources dominated by protostellar envelope emission) and Class II (sources dominated by protoplanetary disk) are segregated based on model IRAC colours.  
The [3.6] -- [4.5] vs [5.8] -- [8.0] IRAC CCP is used and is shown in 
Figure \ref{irac_yso}(a). The boxes which are adopted from \citet{2007A&A...463..175V} demarcate
the location of Class I and Class II sources in the CCP.

\item This identification scheme is based on the slope of the spectral energy distribution.
As explained in \citet{1987IAUS..115....1L}, IRAC spectral index ${\rm (\alpha = d\ log(\lambda F_{\lambda}) / d\ log(\lambda))}$ is calculated for each source using a linear regression fit. Subsequently, we follow the classification scheme of \citet{2008ApJ...682..445C} and
identify Class I $(\alpha > 0)$ and Class II sources $(-2\leqslant \alpha \leqslant 0)$. The number distribution of Class I and II sources are shown in Figure \ref{irac_yso}(b). 

\item NIR CCP is also an efficient tool to understand the nature of stellar populations 
\citep{{2002ApJ...565L..25S},{2004ApJ...608..797O},{2004ApJ...616.1042O},{2006A&A...452..203T}}. As seen in Figure \ref{CC_CM_plot}, the CCP is divided into
three regions. Sources in the `F' region are mainly field stars, Class III or Class II sources with small NIR excess. The `T' region is mainly for classical T-Tauri or Class II stars and the `P' region is populated by Class I sources or reddened
Herbig AeBe stars. From the sources in our sample, we estimate a mean photometric
error of $\sim$ 0.06 mag in all three bands. To account for this photometric uncertainty and the error involved in defining the reddening vector, we take a conservative offset of three times the photometric error to eliminate contamination to the Class II sample
from the field star population. This is shown as the dotted line. It should be 
noted here that there might be a few genuine Class II sources which get filtered
out in this process.  
\end{enumerate}

Based on the above three criteria, we identify 12 Class I, and 80 Class II sources in the probed region and Figure \ref{yso_excitng_dist}(a) shows their distribution
on the MIR 8.0~$\rm \mu m$ IRAC image. The distribution is random with a marginal
overdensity of Class I sources within the dust clump associated with G346.077$-$0.056
and the absence of Class I sources in the region related to G346.056$-$0.021. 
Figures \ref{yso_excitng_dist}(b) and (c) zoom in to the respective \hii\ regions.   
Spectroscopic studies are essential to ascertain the nature of these identified YSOs
and their association with the \hii\ regions. It should be kept in mind that our
YSO sample is not complete. The overwhelming MIR diffuse emission associated with the \hii\ regions, especially G346.077$-$0.056, renders the
detection and photometry of point sources impossible which reflects as lack
of point sources in the GLIMPSE catalog. 

\subsection{Emission from dust component} \label{dust}
\begin{figure*}[t]
\centering
\includegraphics[scale=0.11]{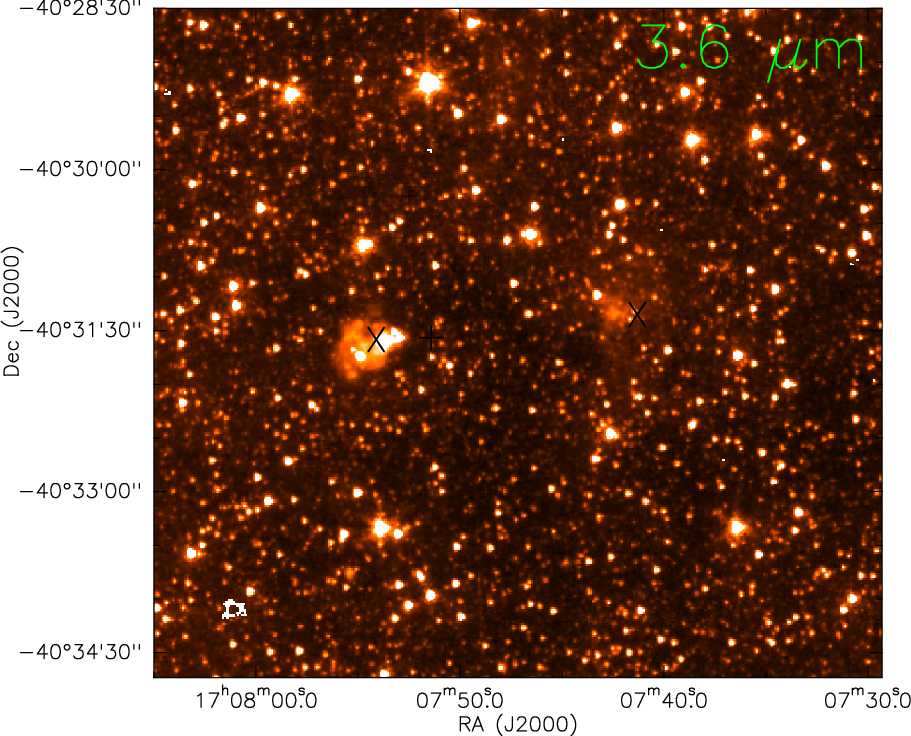}
\includegraphics[scale=0.11]{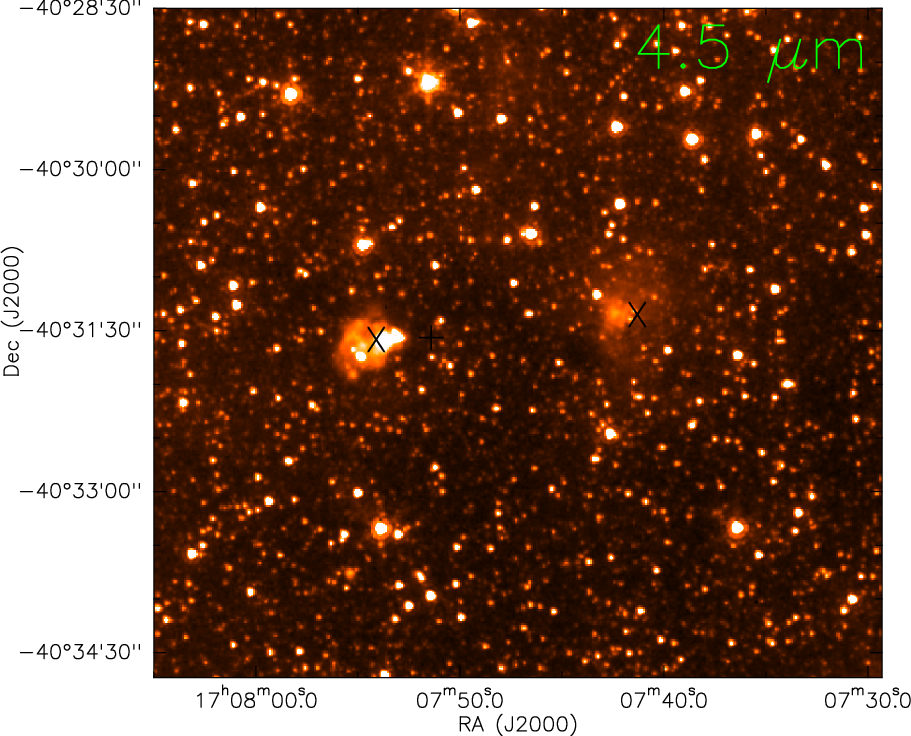}
\includegraphics[scale=0.11]{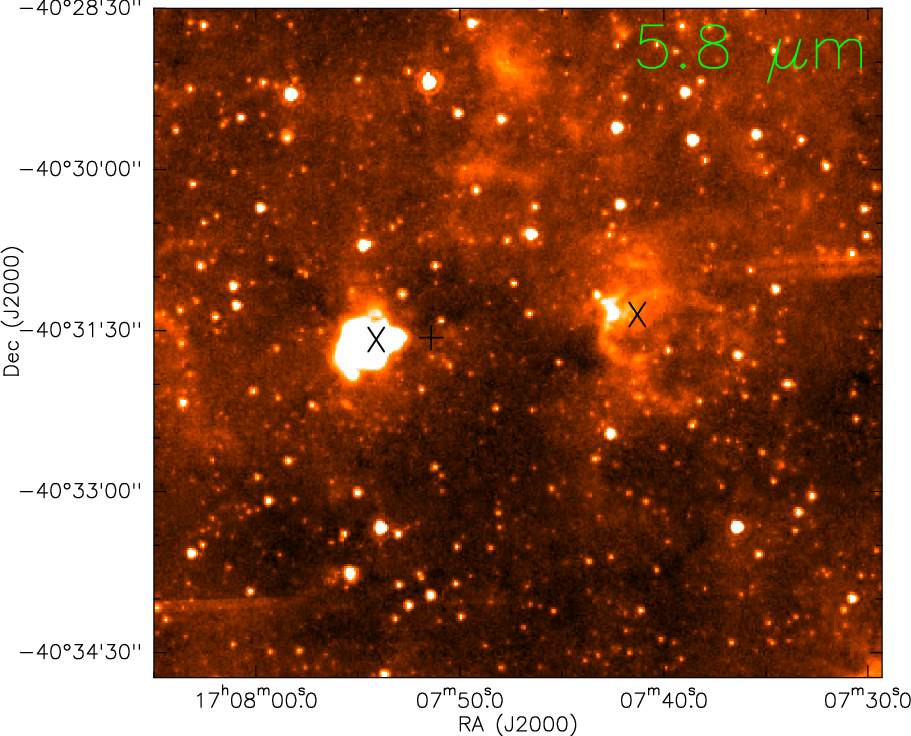}
\includegraphics[scale=0.11]{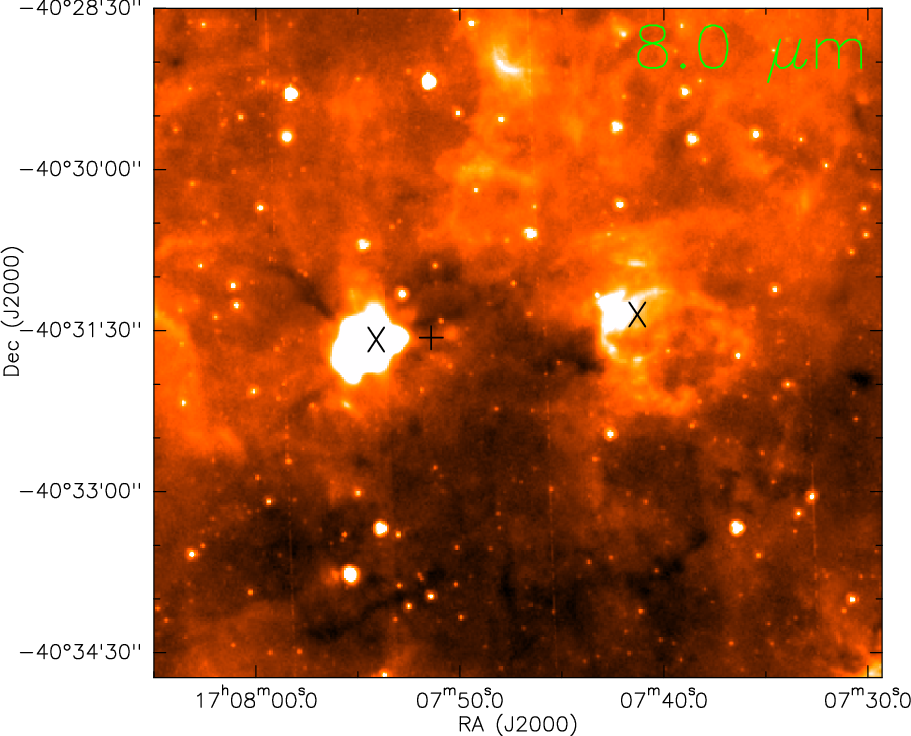}
\includegraphics[scale=0.11]{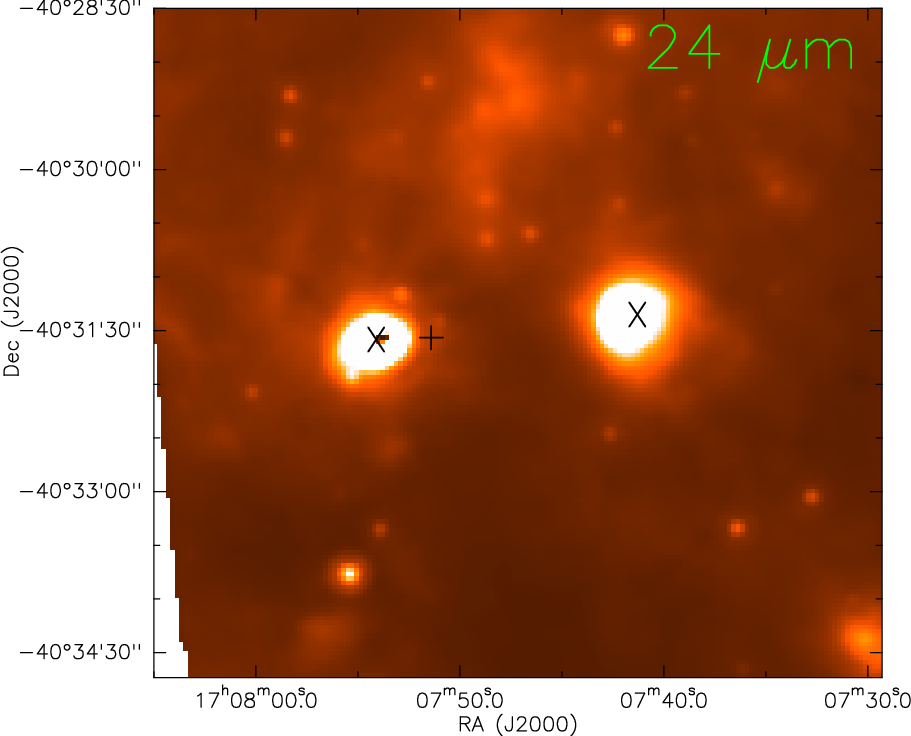}
\includegraphics[scale=0.11]{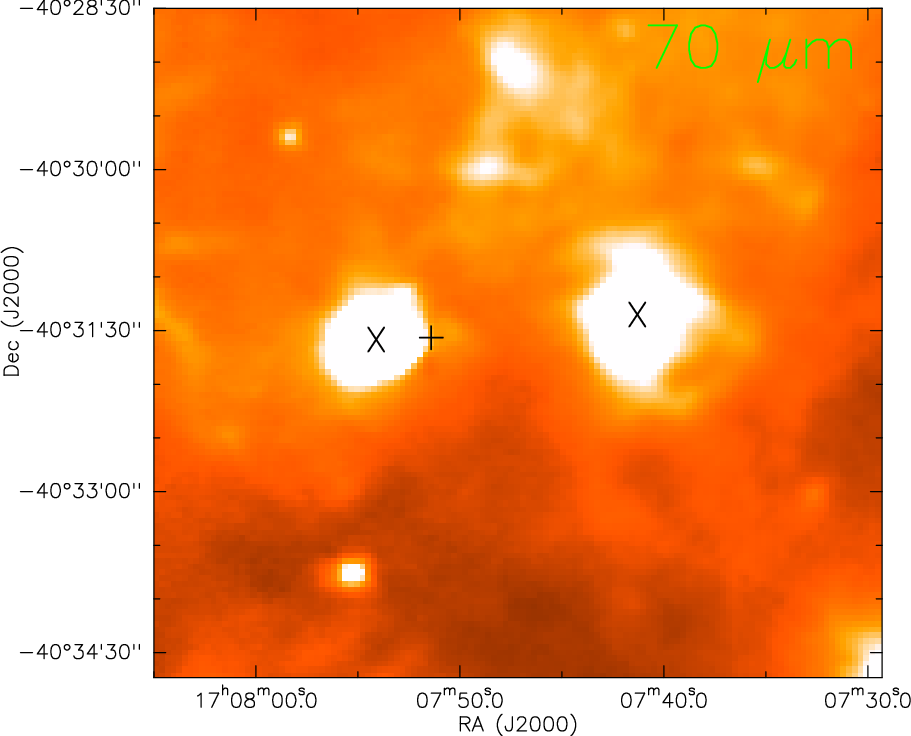}
\includegraphics[scale=0.11]{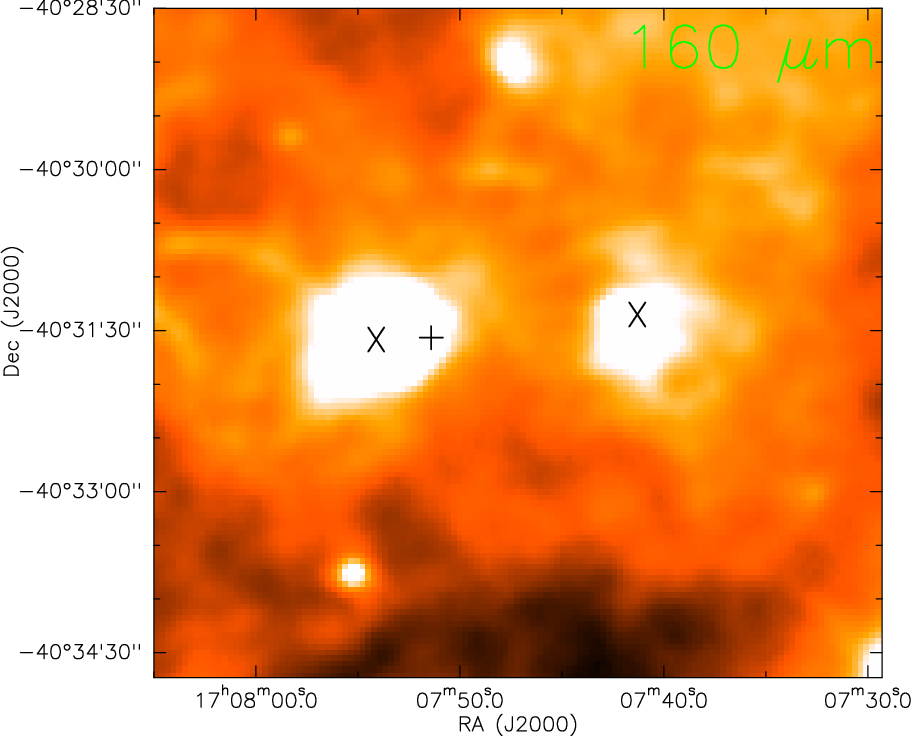}
\includegraphics[scale=0.11]{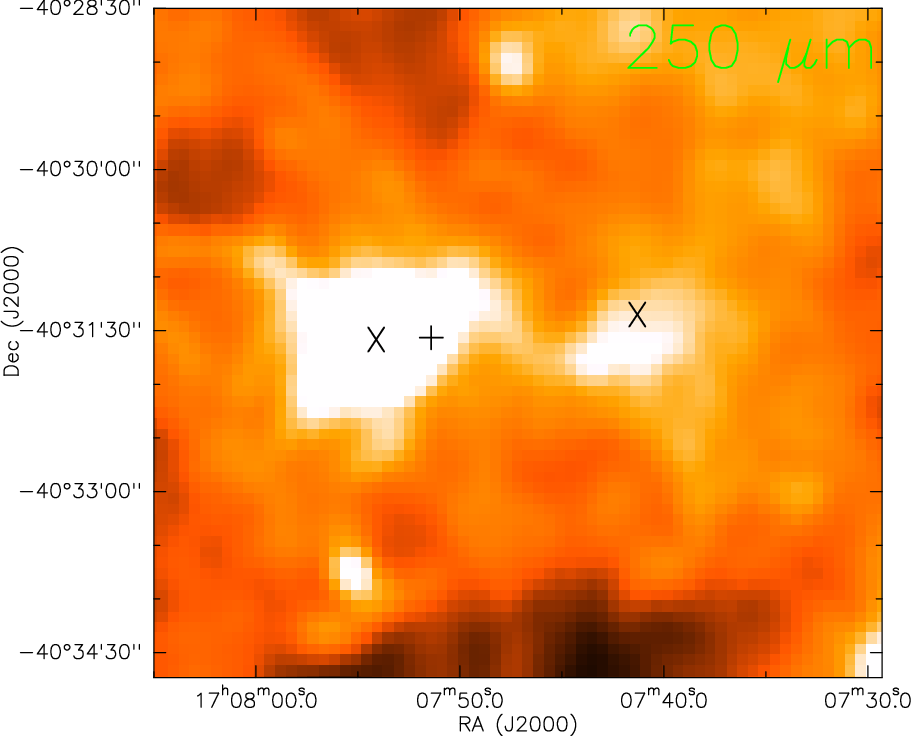}
\includegraphics[scale=0.11]{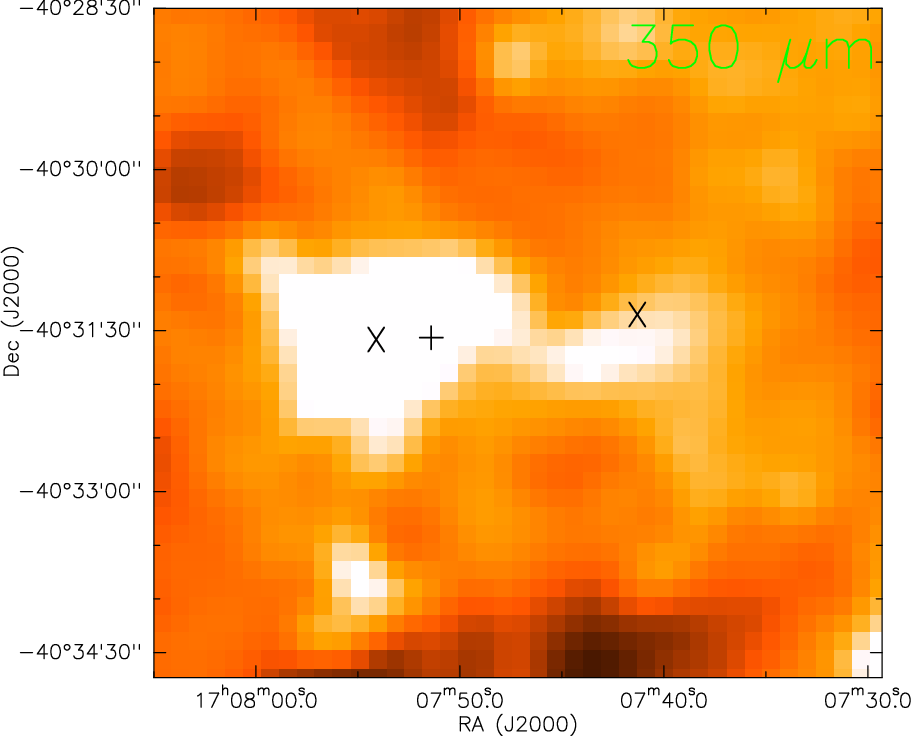}
\includegraphics[scale=0.11]{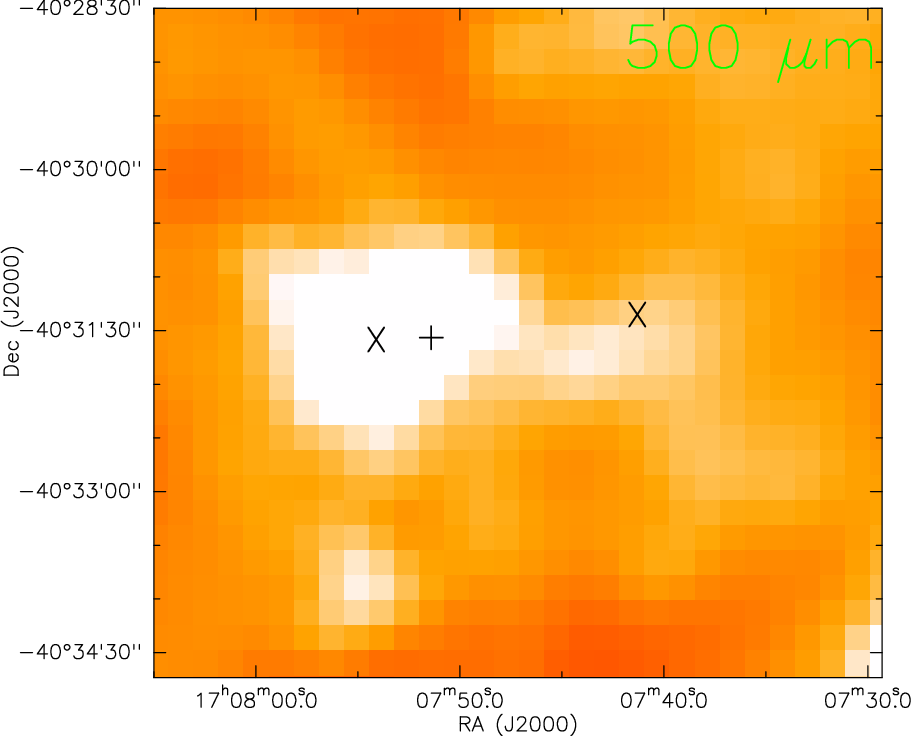}
\caption {Dust emission associated with the \hii\ regions are shown. Top panel from left 3.6, 4.5, 5.8, 8.0, 24~$\rm \mu m$. Bottom panel from left 70, 160, 250, 350, 500~$\rm \mu m$. The `$\times$' marks are the positions of \hii\ positions. The `+' mark show the position of the associated IRAS point source, IRAS 17043$-$4027.}
\label{dust_warm_cold}%
\end{figure*}

Figure \ref{dust_warm_cold} shows the MIR and FIR emission sampled in the IRAC, 
MIPSGAL and Hi-Gal images. Diffuse emission is seen towards the \hii\ regions in 
all the four IRAC bands. 
As discussed in \citet{2008ApJ...681.1341W}, 
various processes contribute to the emission in each
IRAC band. These include thermal emission from the circumstellar dust heated by the 
stellar radiation and emission due to excitation of polycyclic aromatic hydrocarbons 
(PAHs) by UV photons in the Photo Dissociation Regions. In \hii\ regions there would be significant contribution from trapped Ly$\alpha$ heated dust as well \citep{1991MNRAS.251..584H}. Apart from this, diffuse emission in the Br$
\alpha$ and Pf$\beta$ lines, and $\rm H_2$ line emission from shocked gas would
also exist \citep{2008ApJ...681.1341W}.
The shorter IRAC bands (3.6, 4.5~$\rm \mu m$) reveal the point sources 
since emission here is dominated by the stellar photosphere. As seen in the figure, the 
two \hii\ regions show up as faint, and compact nebulosities. The extent and brightness of the diffuse emission
increases from 3.6~$\rm \mu m$ to 8.0~$\rm \mu m$. Further, at 5.8 and 8.0~$\rm \mu m$ G346.056$-$0.021 shows an extended bubble-type morphology towards the south-west.
G346.077$-$0.056 shows an extended, diffuse morphology with an irregular
distribution of MIR emission (see Figure \ref{rad_8mic}). This
will be discussed more detail in a later section.
The comparison of
the [5.8] band, which is mostly a dust tracer, and the [4.5] band, which has
significant contribution from Br$\alpha$ emission ($\rm H^{+}$ tracer) \citep{2004ApJS..154..322C}
shows that
the two \hii\ regions are dominated by dust emission.
The 24~$\rm \mu m$ emission is in unison with the radio free-free emission. 
This emission which spatially correlates well with ionized component 
can be attributed to Ly$\alpha$ heating of normal size dust grains that
could maintain the temperatures close to 100~K in the ionized region \citep{1991MNRAS.251..584H}. Few authors also associate the 24~$\rm \mu m$ emission in \hii\ regions with 
either Very Small Grains or Big Grain replenishment \citep{{2010ApJ...713..592E},{2012ApJ...760..149P}}. As we move towards the
cold dust sampled with the {\it Herschel} bands, filamentary structures are prominent and seem to connect the two \hii\ regions. The extent and brightness of G346.056$-$0.021 decreases 
and that of G346.077$-$0.056 increases as we go longward of 160~$\rm \mu m$ suggesting the dominance of warm dust in G346.056$-$0.021 and cold dust in G346.077$-$0.056.

We study the nature of the cold dust emission using the {\it Herschel} FIR bands.
Line-of-sight average molecular hydrogen column density and dust temperature maps are generated by pixel-wise modified single temperature blackbody fits using the background-corrected fluxes  and assuming the emission at these wavelengths to be optically thin. As discussed and adopted in
several studies \citep{{2010A&A...518L..98P},{2010A&A...518L..99A},{2011A&A...535A.128B},{2016ApJ...818...95L}, {2012A&A...542A..10A}}, we have also excluded the 70~$\rm \mu m$ 
emission. This is because the
emission at 70~$\rm \mu m$ may not be optically thin. Apart from this, 
there would be significant contribution from warm dust components and
hence the SED cannot be modeled with a single
temperature gray body. Thus we have only four points mostly on the Raleigh-Jeans part to constrain the model.

The first step involves converting the SPIRE map units from MJy sr$^{-1}$ to 
Jy pixel$^{-1}$ which is the unit of the PACS images. Subsequent to this, the 
70, 160, 250, 350~$\rm \mu m$ images are convolved and regridded to the lowest resolution (35.7$\arcsec$) and largest pixel size (14$\arcsec$) of the 
500~$\rm \mu m$ image. The convolution kernels are taken from \citet{2011PASP..123.1218A}. The above steps are carried out using the {\it Herschel}
data reduction software HIPE\footnote{The software package for Herschel Interactive Processing Environment (HIPE) is the application that allows users to work with the Herschel data, including finding the data products, interactive analysis, plotting of data, and data manipulation.}.

We have estimated the background flux, $\rm I_{bg}$, in each band from a relatively smooth and dark region devoid of bright diffuse emission and filamentary structures.
This region is located at an angular distance of $\sim1^\circ$ from the \hii\ regions and centered at $\alpha_{2000}$=17:11:57.17, $\delta_{2000}$=$-$41:06:38.99. 
The background value was estimated by fitting a Gaussian function to the 
distribution of individual pixels in the specified region. The fitting was carried 
out iteratively by rejecting the pixels having values outside $\pm$2$\sigma$ till the fit converged \citep{{2011A&A...535A.128B},{2013A&A...551A..98L}}. We have used the same region for the determination of background offset in all the bands. $\rm I_{bg}$ is
estimated to be $-$3.22, 1.45, 0.72, 0.26 Jy pixel$^{-1}$ at 160, 250, 350, and 500~$\rm \mu m$, respectively. The negative flux value at
160~$\rm \mu m$ is due to the arbitrary scaling of the PACS images.

\begin{figure*}
\centering
\includegraphics[scale=0.25]{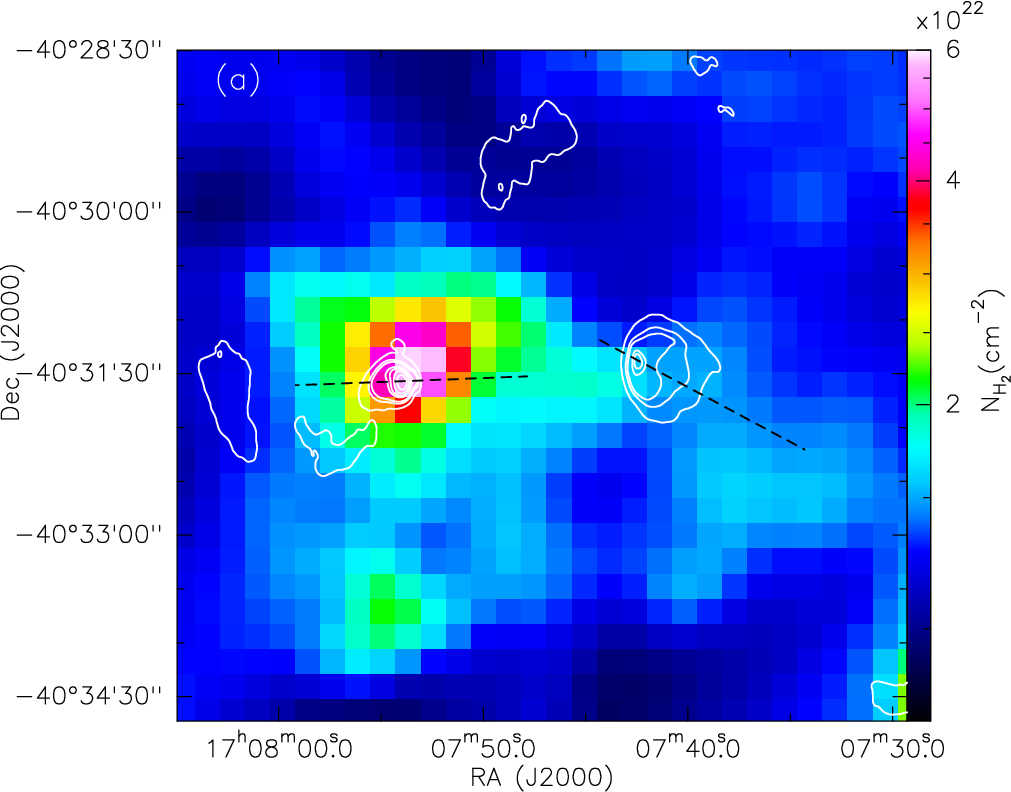}
\includegraphics[scale=0.25]{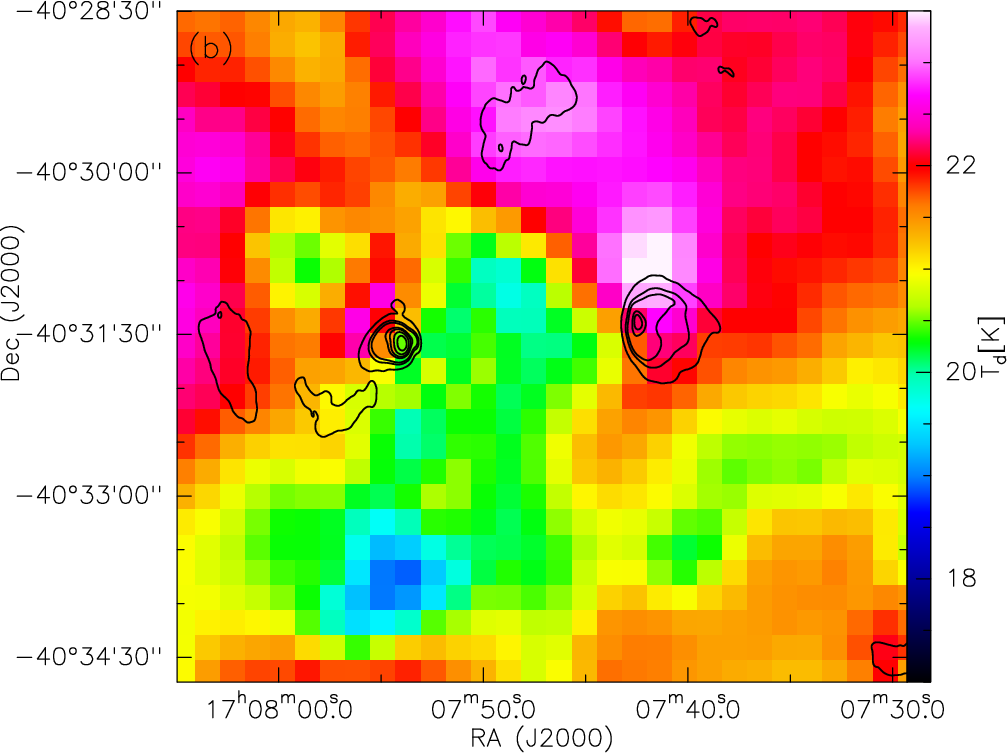}
\includegraphics[scale=0.25]{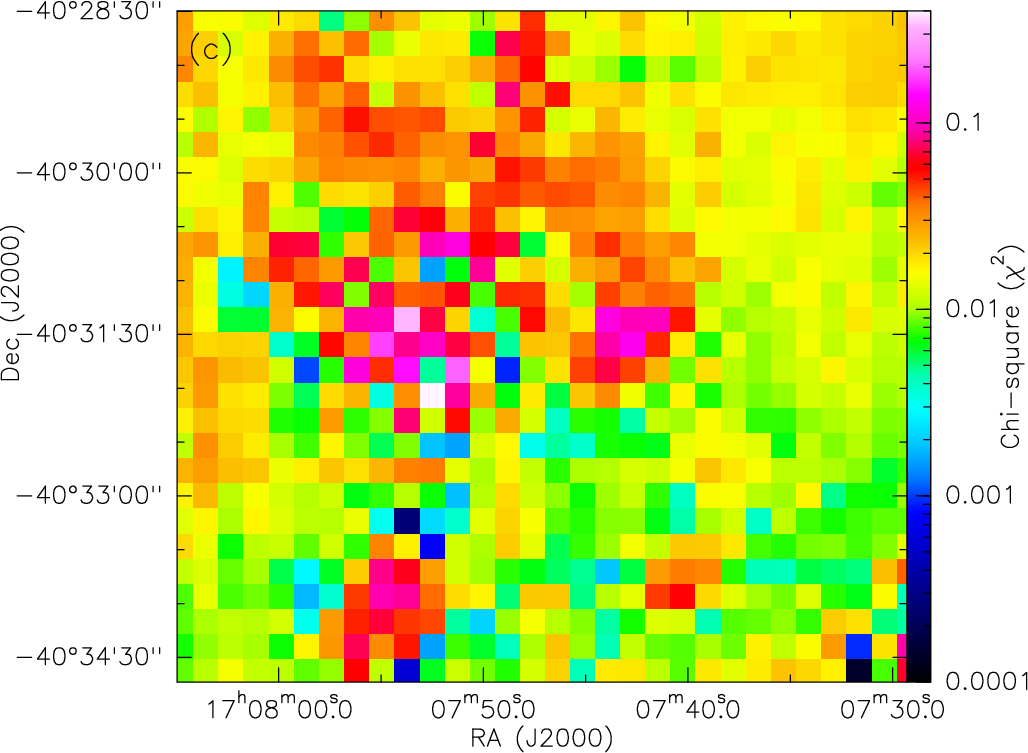}
\caption {Column density (a), dust temperature (b) and chi-square ($\chi^2$) (c) maps of the region associated to \hii\ regions. 1280~MHz GMRT radio emission is shown as contours. The contour levels are same as those plotted in Figure \ref{yso_excitng_dist}. The dashed line on the \hii\ regions shows the projections, the column density variation along which is used to understand the morphology of the ionized region in a later section.}
\label{dust_temp_B2}%
\end{figure*}

The pixel-to-pixel SED fitting was carried out by adopting the 
following expressions \citep{{2011A&A...535A.128B},{2012MNRAS.426..402F},{2013A&A...551A..98L},{2015MNRAS.447.2307M}}. 
\begin{equation}
\rm S_{\nu}(\nu) - I_{bg}(\nu) = B_{\nu}(\nu,T_{d})\ \Omega\ (1-e^{-\tau_{\nu}})
\end{equation}
where 
\begin{equation}
\rm \tau_{\nu} = \mu_{\rm H_{2}}\ m_{\rm H}\ \kappa_{\nu}\ N({\rm H_{2}}) 
\end{equation}
where, $\rm S_{\nu}(\nu) $ is the observed flux density, $\rm I_{bg}(\nu) $ is the 
background flux, $\rm B_{\nu}(\nu,T_{d}) $ is the Planck's function, $\rm T _{d} $ 
is the dust temperature, $\rm \Omega $ is the solid angle (in steradians) from where 
the flux is obtained (solid angle subtended by a 14$'' \times $ 14$''$ pixel), $\rm 
\mu_{\rm H_{2}}$ is the mean molecular weight, $\rm m_{\rm H}$ is the mass of hydrogen atom,
$\rm \kappa_{\nu}$ is the dust opacity and $\rm N$(H$_{2} $) is the column density. 
We have assumed a value of 2.8 for  $\mu_{\rm H_{2}}$ \citep{2008A&A...487..993K}. 
The dust opacity $\rm \kappa_{\nu} $ 
is defined to be $\rm \kappa_{\nu} = 0.1~(\nu/1000~{\rm GHz})^{\beta}~{\rm cm^{2}/g} 
$, where, $\beta$ is the dust emissivity spectral index \citep{{1983QJRAS..24..267H}, {1990AJ.....99..924B}, {2010A&A...518L.102A}}. The above model was
fitted to the four data points using the non-linear least square Levenberg-Marquardt algorithm, where $\rm 
T_{d}$ and $\rm N$(H$_{2})$ are kept as free parameters. Given the limited number of data points, we prefer to fix the value of $\beta$ to 2
\citep{{1983QJRAS..24..267H}, {1990AJ.....99..924B}, {2010A&A...518L.102A}} which is also a typical value estimated for a large sample of \hii\ regions \citep{2012A&A...542A..10A}.
We have used a 
conservative 15\% uncertainty on the background subtracted flux densities 
\citep{2013A&A...551A..98L}. The generated column density, dust temperature  
maps  alongwith the corresponding $
\chi^2$ map are shown in Figure \ref{dust_temp_B2}. We have overlaid the 1280~MHz radio map to correlate the emission from 
ionized gas and the cold dust component. The fitting uncertainties are small as
is evident from the $\chi^2$ map where the maximum value for individual pixel fits
is $\sim$ 1.  
    
The column density map shows a large clump enveloping the \hii\ region G346.077$-$0.056 with the previously mentioned western filamentary structure visible.
The region associated with G346.056$-$0.021 shows relatively low column density which is indicative of a less significant cold dust component. The
peak (5.9$\times 10^{22}\rm\ cm^{-2}$) is located close to the radio peak of
G346.077$-$0.056. Apart from the \hii\ regions, a compact, spherical clump is seen to the south of G346.077$-$0.056. The association of this clump with the \hii\ regions
is not certain as there is no literature available on it. 
As expected, we see regions of higher dust temperature towards both the \hii\ regions. The dust temperature map shows extended warm distribution towards
the north-east of G346.056$-$0.021 with the dust temperature distribution peaking
just north of G346.056$-$0.021. This extended distribution coincides with a faint diffuse ionized structure (detected at the 3$\sigma$ level
of the 1280~MHz emission). The southern compact clump displays the coldest 
dust temperature.
 
\begin{figure}
\centering
\includegraphics[scale=0.3]{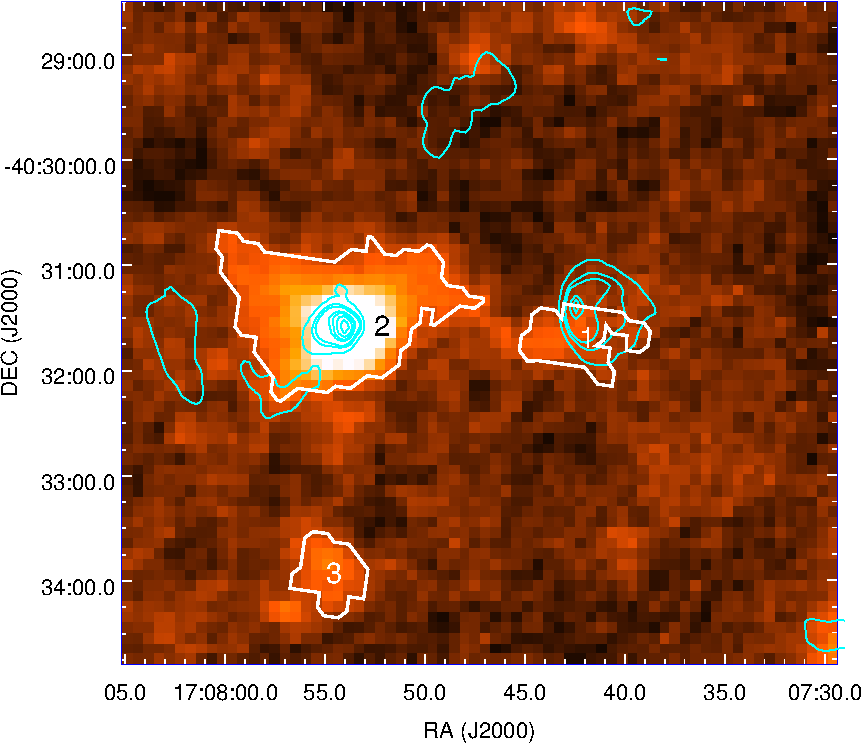}
\caption {ATLASGAL image shown, on top of which the clump apertures are overlaid. The clumps are labeled as 1, 2 and 3. 1280~MHz GMRT radio emission is shown as cyan contours with the levels same as those plotted in Figure \ref{yso_excitng_dist}.}
\label{clumps}%
\end{figure}

To identify and study the cold dust clumps, we use the 
ATLASGAL 870~$\rm \mu m$ map because this wavelength is 
sensitive to the colder dust components and also the emission is optically thin. 
Further, the resolution of the ATLASGAL map is better (18\arcsec.2) compared to
the column density map (35\arcsec.7). 
We use the 2D {\it Clumpfind} algorithm \citep{1994ApJ...428..693W} with a 2$\sigma$ (where, $\sigma$ = 0.06~Jy/beam) threshold
and optimum contour levels to detect the clumps. Using this threshold three clumps are detected, the retrieved apertures of which are shown overlaid on the ATLASGAL image in Figure \ref{clumps}. Clump 1 overlaps with the southern part of the \hii\ region G346.056$-$0.021  and is mostly part of the filamentary structure.
Clump 2 is seen to be associated and enveloping G346.077$-$0.056. Clump 3 is
located towards the south of G346.077$-$0.056.
Masses of the clumps are estimated using the following expression
\begin{equation}
\rm M_{\rm clump} = \mu_{\rm H_{2}} m_{\rm H} A_{\rm pixel} \Sigma N ({\rm H_{2}}) 
\label{clump_mass1}
\end{equation}
where, $\rm m_{\rm H} $ is the mass of hydrogen, $\rm A_{\rm pixel} $ is the pixel 
area in $\rm cm ^{2} $, $ \mu_{\rm H_{2}} $ is the mean molecular weight and $\rm 
\Sigma N ({\rm H_{2}})$ is the integrated column density over the pixel area. Clump 
apertures retrieved from the {\it Clumpfind} algorithm are used to determine $\rm 
\Sigma N ({\rm H_{2}})$ from the column density map. Location of 870~$\rm \mu m$ peaks, deconvolved sizes, mean dust temperature, mean column density, integrated column density, masses and   
number density ($\rm n_{H_2}= 3\Sigma \,NH_2\,/4r$, $r$ being the radius) of the clumps are estimated and listed in Table \ref{clump_properties}. 
Clump 2 has been studied by \citet{2015MNRAS.446.2566Y}. They have estimated the
mass of the clump from 870~$\rm \mu m$ integrated flux density. Assuming a 
dust temperature of 30~K, they obtain a mass of 13013 M$_{\odot}$.
This is a factor of $\sim$ 1.2 lower than the estimates from our column density
map. The possible reason for the higher mass estimate in our work could be
the lower dust temperature (21~K) and different dust opacity assumed. One also
cannot exclude the effect of flux loss associated with ground based observations \citep{2017A&A...602A..95L}.

\begin{table*}[h]
\centering
\caption{Physical parameters of the clumps. The columns refer to location of 870~$\rm \mu m$ peaks, deconvolved 
sizes, mean dust temperature, mean column density, 
integrated column density, mass and   
number density ($\rm n_{H_2}= 3\Sigma \,NH_2\,/4r$)}
\label{clump_properties}
\tiny
\begin{tabular}{ccccccccc}
\\ \hline
Clump No. &\begin{tabular}[c]{@{}c@{}}RA (2000)\\
(hh:mm:ss.ss)\end{tabular} & \begin{tabular}[c]{@{}c@{}}DEC (2000)\\ (dd:mm:ss.ss)\end{tabular} & 
\begin{tabular}[c]{@{}c@{}}Radius \\ (pc)\end{tabular} & \begin{tabular}[c]{@{}c@{}}Mean $T_{d}$\\ (K)\end{tabular} & 
\begin{tabular}[c]{@{}c@{}}Mean $N(\rm H_{2})$\\ 
($\rm \times 10^{22}cm^{-2}$)\end{tabular} &  
\begin{tabular}[c]{@{}c@{}}$\sum N(\rm H_{2}$)\\ 
($\rm \times 10^{23}cm^{-2}$)\end{tabular}  & \begin{tabular}[c]{@{}c@{}}Mass\\ (M$_{\odot}$)\end{tabular} &  
\begin{tabular}[c]{@{}c@{}}Number density ($\rm n_{H_2}$)\\ 
($\rm \times 10^4 cm^{-3}$)\end{tabular} \\
\hline
1 & 17:07:43.32 & $-$40:31:47.77  & 0.6 & 21.6  & 1.5  & 1.6  & 1922 & 3.1 \\
2 & 17:07:12.02 & $-$40:36:57.00  & 1.3 & 21.0  & 2.9  & 12.5 & 15248 & 2.4 \\
3 & 17:07:09.40 & $-$40:37:09.09  & 0.2 & 19.1  & 1.9  & 1.2  & 1412 & 45.9 \\                                                  
\hline
\end{tabular}
\end{table*}

\section{Morphology of the \hii\ regions}
\label{morph}
\begin{figure*}[t]
\centering
\includegraphics[scale=0.25]{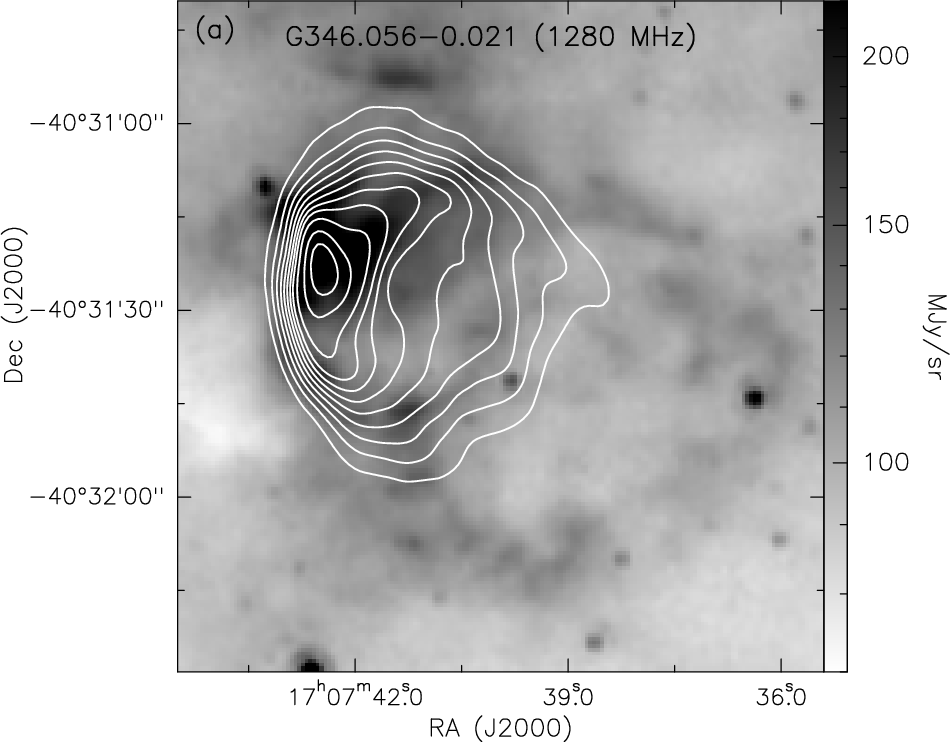}
\includegraphics[scale=0.25]{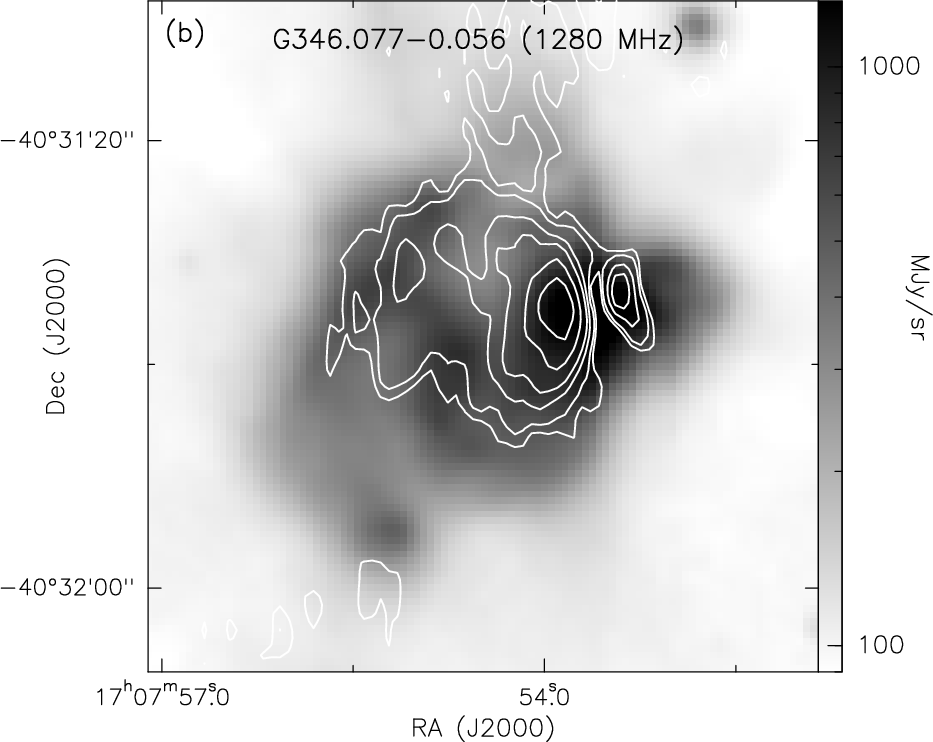}
\caption{1280~MHz radio contours overlaid on the 8.0~$\rm \mu m$ grey scale image of (a) G346.056$-$0.021 and (b) G346.077$-$0.056.  In (a) we have 
overlaid the lower resolution contours presented in Figure \ref{radio_maps_r1}
and in (b) to show the finer structures, we plot the higher resolution contours
as shown in Figure \ref{radio_maps_rm5}.}
\label{rad_8mic}%
\end{figure*}

As mentioned in Section \ref{ionized}, both the \hii\ regions show signature of
cometary morphology in the radio. This morphology shows up as a bright, arc-type head and a 
diffuse, broad tail emission.  
In Figure \ref{rad_8mic}, we compare the radio morphology with the MIR emission
in IRAC 8.0~$\rm \mu m$ band. The radio and MIR emission associated with 
G346.056$-$0.021 is seen to spatially correlate towards the head but there is
a void in the MIR emission towards the tail implying that the \hii\ region is
density-bounded towards the south-west. 
The picture presented by G346.056$-$0.021 is similar to that of the cometary \hii\ region G331.1465+00.1343 shown in Fig. 1 of \citet{2007prpl.conf..181H}. In contrast, the MIR emission
associated with G346.077$-$0.056 is extended and envelops most of the ionized region.
The radio peak and the possible second \hii\ region is seen to coincide with
the enhanced MIR emission towards the east.

In this section, we attempt to understand the origin of such cometary morphologies that are commonly seen in \hii\ regions \citep{1989ApJS...69..831W}.
Several models have been proposed in literature to address this of which the 
frequently used ones are (i) the bow-shock model and (ii) the champagne flow model.
The former is due to the
interaction of a supersonically moving, wind-blowing, ionizing star with the dense 
surrounding molecular gas. The latter model is a result of a steep 
density gradient encountered by an expanding \hii\ region around a nearly stationary, 
ionizing source. Here, the ionizing star is possibly located at the edge of a
dense clump where the ionized gas expands asymmetrically out towards a regions of 
minimum density.
Initial papers by \citet{{1985ApJ...288L..17R},{1990ApJ...353..570V},{1991ApJ...369..395M},{1978A&A....70..769I},{1979A&A....71...59T}} offer a detailed
insight into these two models and the related physical conditions favouring
one over the other.
Recently, more involved radiation-hydrodynamic models have been proposed 
that include 
density gradient in the surrounding medium, stellar wind contribution from
ionizing star apart from its supersonic motion within the ambient cloud \citep{2006ApJS..165..283A}. Another recent paper by \citet{2014MNRAS.438.1335R} explores  a  quasi-one-dimensional,  
steady-state wind model to explain the cometary morphology of UC\hii\ regions.
High spatial and spectral resolution observations of the MIR [Ne II] fine structure
line in a sample of \hii\ and UC\hii\ regions by \citet{{2005ApJ...631..381Z},{2008ApJS..177..584Z}} have
shed further light on the various models. These studies show the co-existence
of \hii\ regions with dense, and massive molecular cores where the ionized
emission display parabolic shells with the open end facing away from the dense cores
and the kinematics reveal gas flow tangential to these shells. RRL studies of
cometary \hii\ regions have also invoked `hybrid' (combination
of bow-shock and champagne flow) models to explain the observed velocity 
structure \citep{2014A&A...563A..39I}.

Based on simple analytic expressions and arguments, we aim at interpreting the
observed morphology of the ionized emission and the column density distribution associated with our \hii\ regions.
We first consider the bow-shock model and derive the bow-shock parameters along
the lines discussed in \citet{{1985ApJ...288L..17R},{1990ApJ...353..570V},{1991ApJ...369..395M}}. In this model, the stellar wind freely streams
in all direction until it encounters a terminal shock. 
In the direction of motion, the shock occurs at a distance $\it l_s$ (stand-off-distance) from the star, where the momentum flux of the wind equals ram pressure of the surrounding ambient ISM. This `bow-shock' gives rise to an \hii\ region resembling a thin paraboloidal shell in the plane of the sky. 

The stand-off-distance  $\it l_s$ is estimated using the following expressions \citep{{1990ApJ...353..570V},{1991ApJ...369..395M}}  

\begin{equation}
\rm {\it l_s} = 5.50\times10^{16}\ \dot{m}_{*,-6}^{1/2}\ v_{w,8}^{1/2}\ \mu_{H}^{-1/2}\ n_{H,5}^{-1/2}\ v_{*,6}^{-1}\ cm
\label{stand-off_dis}
\end{equation} 
\begin{equation}
\rm \dot{m}_{*} =  2\times10^{-7}\ (L/L_{\odot})^{1.25} \ M_{\odot}\ yr^{-1} 
\end{equation}
\begin{equation}
\rm logv_{w} = -38.2\ + 16.23\ logT_{eff}\ -1.70\ (logT_{eff})^{2} 
\end{equation}
where, $\rm \dot{m}_{*,-6}$ = $\rm \dot{m}_{*}\times10^{6}\ M_{\odot}\ yr^{-1}$ is 
the stellar wind mass loss rate, $\rm v_{w,8}$ = $\rm v_{w}\times10^8 cm\ sec^{-1}$ is the terminal velocity of wind, $\rm \mu_{H}$ is the mean mass per hydrogen nucleus, $\rm n_{H,5}=n_{H}\times10^5\ cm^{-3}$ is the hydrogen gas density and $\rm v_{*,6} = \rm v_{*}\times10^6\ cm\ sec^{-1}$ is the relative velocity of star. L and $\rm L_{\odot}$ are the luminosity of star and sun respectively. $\rm T_{eff}$ is the effective temperature of the star.
The medium through which the star moves would be a combination of ionized and  
neutral medium. We consider a neutral medium here and take $\rm \mu_{H} = \rm 1.4~m_{H}$, where $\rm m_H$ is taken as one atomic mass unit. 
We estimate $\rm n_{H}$ to be $\rm 6.2 \times 10^4\  cm^{-3}$ and $\rm 4.8 \times 10^4\ cm^{-3}$ for G346.056$-$0.021 and G346.077$-$0.056, respectively from the column density maps (refer Section \ref{dust}). 
For the estimated spectral type of the ionizing stars, we assume luminosities in
the range $1.0\times10^5$ - $1.3\times10^5 \rm L_{\odot}$ and $6.6\times10^4$ - $8.0\times10^4 \rm L_{\odot}$ for G346.056$-$0.021 and G346.077$-$0.056, respectively from \citet{2005A&A...436.1049M}. We further assume a typical velocity of $\rm 10~km\ sec^{-1}$ \citep{1992ApJ...394..534V} for the ionizing star. 
Plugging in these values in the above equations, we calculate the expected stand-off distances. For G346.056$-$0.021 we obtain a value of 0.4$\arcsec$ (0.02~pc) - 0.5$\arcsec$ (0.03~pc) and for G346.077$-$0.056
we estimate 0.2$\arcsec$ (0.01~pc) - 0.3$\arcsec$ (0.016~pc).
In the image plane, the stand-off distance is defined as the distance between the
steep density gradient at the cometary head and the radio peak (assuming 
this to be the location of ionizing star). Hence from the the radio maps, we estimate $l_s$ to be $\sim 9\arcsec$ ($\rm 0.5~pc$) and $\sim 11\arcsec$ ($\rm 0.6~pc$) for G346.056$-$0.021 and G346.077$-$0.056, respectively. These values are significantly larger
than the expected theoretical values. The discrepancies between the theoretical
and observed stand-off distance estimates narrows down if we consider ionizing stars to move slower, with velocity of $\rm \sim 1~km\ sec^{-1}$.
 It should be noted here that the ionizing star 
need not always be at the location of the radio peak \citep{2003A&A...405..175M} and that the viewing angle
would also play a role in the estimated stand-off distance.

Further light can be shed by deriving the trapping parameter ($\rm \tau_{bs}$). 
As discussed in \citet{1991ApJ...369..395M}, the swept up dense shells by the
supersonically moving, ionizing star trap the \hii\ region and the ram pressure inhibits their further dynamical expansion. This parameter is so defined that its
inverse gives the ionization fraction. The shell thus traps the \hii\ region when
there are more recombinations in the shell compared to ionizing photons. This
happens when $\rm \tau_{bs} > 1$. This parameter is shown to be much larger 
($\rm \tau_{bs} >> 1$) as computed by \citet{1991ApJ...369..395M} for a sample of cometary \hii\ regions. We estimate $\rm \tau_{bs}$ for G346.056$-$0.021 and G346.077$-$0.056 using the following expression from \citet{1991ApJ...369..395M}

\begin{equation}
\rm \tau_{bs} = 0.282\ \dot{m}_{*,-6}^{3/2}\ v_{w,8}^{3/2}\ n_{H,5}^{1/2}\ T_{II,4}^{-1}\ N_{ly49}^{-1}\ \mu_{H}^{-1/2}\ \gamma^{-1}\ \alpha_{-13}
\label {trapping_parm}
\end{equation}
where, $\rm T_{II} = \rm T_{II,4}\times10^4$ is the temperature of ionized gas, $\rm N_{ly} = \rm N_{ly49}\times10^{49}$ is the ionizing photon flux, $\rm \gamma$ is the ratio between specific heats, $\alpha$ is the hydrogen recombination rate to all levels but the ground state, given in unit of $10^{-13} \rm cm^3\ sec^{-1}$. 
$\rm N_{ly}$ is $\rm 3.2\times10^{48}\ photons\ sec^{-1}$ and $\rm 1.6\times10^{48}\ photons\ sec^{-1}$ for G346.056$-$0.021 and G346.077$-$0.056, respectively (refer Section \ref{ionized}). The temperature of ionized gas is 5500~K and 8900~K for G346.056$-$0.021 and G346.077$-$0.056, respectively and the value of $\rm \gamma$ to be 5/3. The value of $\alpha$ is estimated to be 4.3 $\times 10^{-13}\rm cm^3\ sec^{-1}$ and 3.0 $\times 10^{-13}\rm cm^3\ sec^{-1}$ at 5500~K and 8900~K, respectively by a linear interpolation of the optically thick case from \citet{1989agna.book.....O}. The derived values of $\rm \tau_{bs}$ lie in the
range 4.3 -- 2.7  and 1.8 -- 1.2 for G346.056$-$0.021 and G346.077$-$0.056, respectively.
These values suggest weak or no bow-shock.

\begin{figure*}
\centering
\includegraphics[scale=0.27]{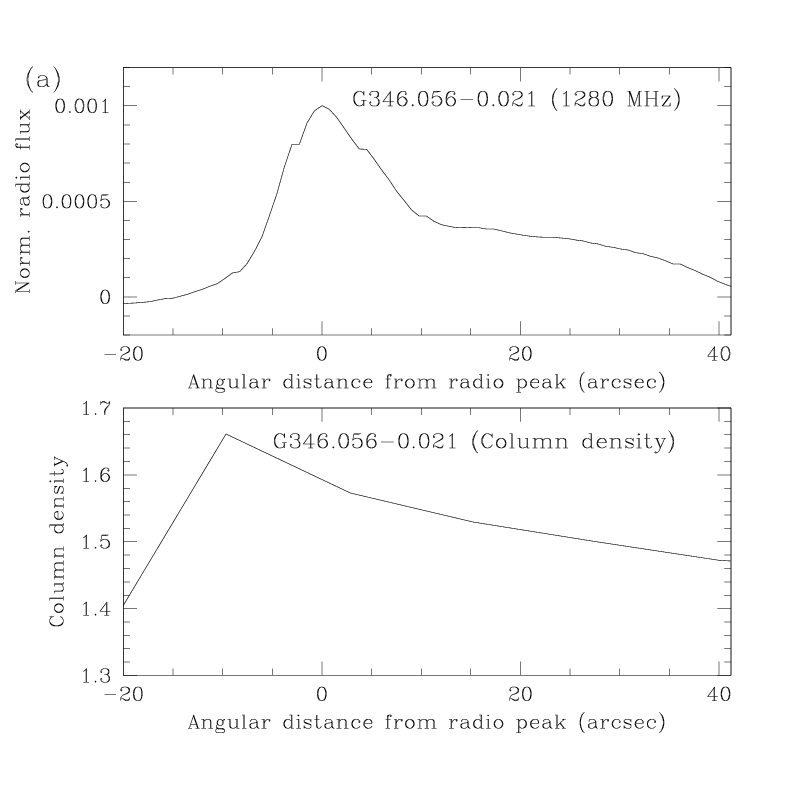}
\includegraphics[scale=0.27]{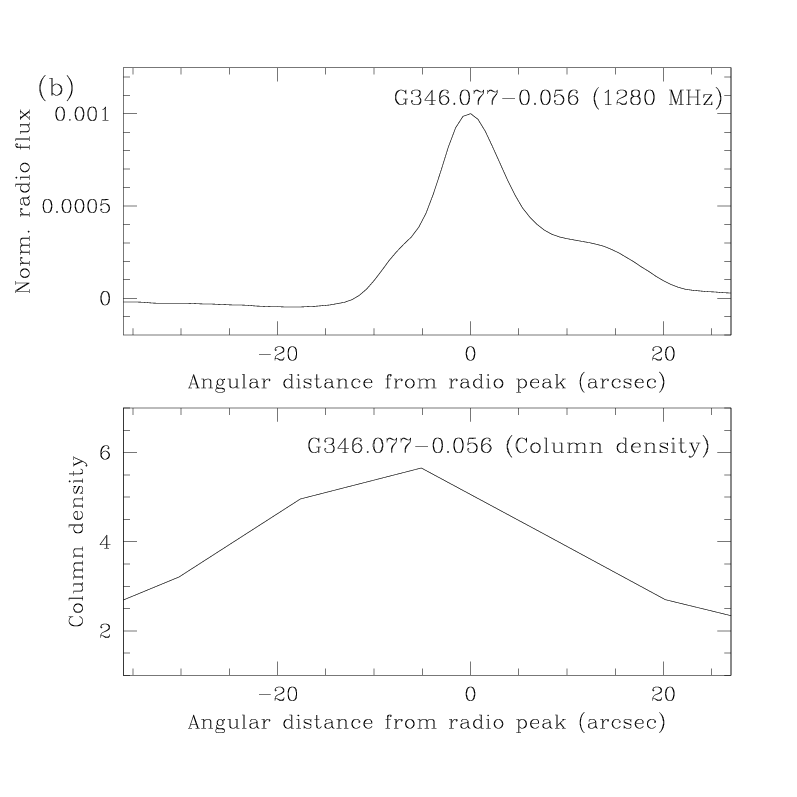}
\caption{Relative distribution of radio emission of the radio emission and column density with respect to the radio peaks for G346.056$-$0.021 and G346.077$-$0.056 along the projections shown in the column density map in Figure \ref{dust_temp_B2}. Zero on the $x$-axis corresponds to the position of the radio peaks increasing towards the direction 
of the tail (south-west for G346.056$-$0.021 and east for G346.077$-$0.056). The radio flux densities
ploted are normalized to the peak flux densities and the column density are given
in terms of $\rm 10^{22}\,cm^{-2}$.}
\label{champ_model}%
\end{figure*}

Considering the above, it is less likely that the bow-shock mechanism is in play
in G346.056$-$0.021 and G346.077$-$0.056. Given this, we explore the the alternate model of champagne
flow. Here, the morphology of the \hii\ region is mostly dictated by the density
structure of the molecular cloud. The \hii\ region expands preferentially
towards low-density regions resulting in a champagne-flow.
To probe this, we try to understand the variation in ionized emission and
correlate with the
column density distribution along the respective cometary axes which are
marked in Figure \ref{dust_temp_B2}. We choose the 1280~MHz map. These projected lines also pass through 
the respective radio peaks of G346.056$-$0.021 and G346.077$-$0.056. The profiles
are displayed in Figure \ref{champ_model}. 
Here, the $x$-axis shows the
positional offset from the radio peak increasing towards the direction of the `tail'. The $y$-axis of the top panels which show the variation in radio flux density is normalized to the peak values. In the bottom panels, the column density
values are given in terms of $\rm 10^{22}\,cm^{-2}$. 
The ionized emission profiles are characteristic of cometary morphology \citep{1989ApJS...69..831W}. What is interesting is the correlation with the density structure. As visible from the plots,
the column density distribution peaks ahead of the ionized emission for 
both the regions implying that dense molecular gas is located at the head of the 
cometary arc of the \hii\ regions which stalls the ionization front thus
keeping the \hii\ regions pressure and ionization bounded in the 
north-east and east directions, respectively for G346.056$-$0.021 and G346.077$-$0.056. 
On the other side, ionized gas streams away into the more rarefied environment
which reveals as a decreasing column density distribution. 
These are signatures of a champagne-flow. Thus it is likely that the cometary
morphology is due to the density gradient rather than the supersonic
motion of the ionizing star. It should be noted here that
the above discussions are based on the projected morphology of
the ionized and molecular gas where we have not considered the
effect of viewing angle. One also has to keep in mind that the resolution and pixel sizes of the
two maps used here are very different and hence crucial small scale correlation is not possible. Though the analytical calculation based on simple assumptions
do not present a strong case for a bow-shock scenario and the observed morphology augurs well with
the champagne-flow model, it is necessary to study in detail the gas kinematics with the highest spectral and spatial resolution in order to 
understand the physical mechanism responsible for the cometary structure of
the ionized regions. 

\section{Summary}
\label{summ}
We have carried out a detailed study of the complex associated with the 
two southern galactic \hii\ regions G346.056$-$0.021 and G346.077$-$0.056. Based on our analysis we deduce the following

\begin{enumerate}
\item Radio continuum emission is detected towards both \hii\ regions at 610
and 1280~MHz. The ionized emission morphology shows a cometary
structure for G346.056$-$0.021. The morphology for G346.077$-$0.056 is mostly compact and spherical with
a faint cometary signature. 

\item The ZAMS spectral type of the ionizing sources are estimated to lie in
the range O7.5V - O7V and O8.5V - O8V for G346.056$-$0.021 and G346.077$-$0.056, respectively. The
dynamical age of G346.056$-$0.021 and G346.077$-$0.056 are similar and estimated to be $\rm \sim 0.5 - 0.2 Myr$.

\item Emission from the dust component shows cold dust to be predominantly located
near G346.077$-$0.056 and the region associated with G346.056$-$0.021 contains relatively warmer dust. The
column density map shows the presence of a dense clump towards G346.077$-$0.056. Two additional
clumps are detected in the 870~$\rm \mu m$ image out of which one is towards G346.056$-$0.021.

\item Assuming the clumps to be physically associated and hence at the same distance, the masses are estimated to be 1922, 15248 and $\rm 1412~ M_{\odot}$ for
clumps 1, 2, and 3, respectively from the column density map. 

\item Simple analytical calculations show that the bow-shock mechanism 
is less likely to be responsible for the observed cometary morphology. The variation of the ionized gas and the column density along the cometary axis favours the 
champagne flow model for both \hii\ regions.

\end{enumerate}

\begin{acknowledgements}
We thank the referee for valuable comments and suggestions, which have helped to improve the quality of the paper.
We thank the staff of the GMRT that made the radio observations possible. GMRT is run by the National Centre
for Radio Astrophysics of the Tata Institute of Fundamental Research. {This publication makes use of 
data products from the Two Micron All Sky Survey, which is a joint project of the University of 
Massachusetts and the Infrared Processing and Analysis Center/California Institute of Technology, 
funded by the NASA and the NSF.} Point Source Catalog (PSC) \citep {2006AJ....131.1163S}
\end{acknowledgements}

\bibliographystyle{aa}
\bibliography{refer.bib}

\begin{thebibliography}{103}
\expandafter\ifx\csname natexlab\endcsname\relax\def\natexlab#1{#1}\fi

\bibitem[{{Allen} {et~al.}(2004){Allen}, {Calvet}, {D'Alessio}, {Merin},
  {Hartmann}, {Megeath}, {Gutermuth}, {Muzerolle}, {Pipher}, {Myers}, \&
  {Fazio}}]{2004ApJS..154..363A}
{Allen}, L.~E., {Calvet}, N., {D'Alessio}, P., {et~al.} 2004, \apjs, 154, 363

\bibitem[{{Anderson} {et~al.}(2011){Anderson}, {Bania}, {Balser}, \&
  {Rood}}]{2011ApJS..194...32A}
{Anderson}, L.~D., {Bania}, T.~M., {Balser}, D.~S., \& {Rood}, R.~T. 2011,
  \apjs, 194, 32

\bibitem[{{Anderson} {et~al.}(2015){Anderson}, {Hough}, {Wenger}, {Bania}, \&
  {Balser}}]{2015ApJ...810...42A}
{Anderson}, L.~D., {Hough}, L.~A., {Wenger}, T.~V., {Bania}, T.~M., \&
  {Balser}, D.~S. 2015, \apj, 810, 42

\bibitem[{{Anderson} {et~al.}(2012){Anderson}, {Zavagno}, {Deharveng},
  {Abergel}, {Motte}, {Andr{\'e}}, {Bernard}, {Bontemps}, {Hennemann}, {Hill},
  {Rod{\'o}n}, {Roussel}, \& {Russeil}}]{2012A&A...542A..10A}
{Anderson}, L.~D., {Zavagno}, A., {Deharveng}, L., {et~al.} 2012, \aap, 542,
  A10

\bibitem[{{Anderson} {et~al.}(2010){Anderson}, {Zavagno}, {Rod{\'o}n},
  {Russeil}, {Abergel}, {Ade}, {Andr{\'e}}, {Arab}, {Baluteau}, {Bernard},
  {Blagrave}, {Bontemps}, {Boulanger}, {Cohen}, {Compi{\`e}gne}, {Cox},
  {Dartois}, {Davis}, {Emery}, {Fulton}, {Gry}, {Habart}, {Huang}, {Joblin},
  {Jones}, {Kirk}, {Lagache}, {Lim}, {Madden}, {Makiwa}, {Martin},
  {Miville-Desch{\^e}nes}, {Molinari}, {Moseley}, {Motte}, {Naylor}, {Okumura},
  {Pinheiro Gon{\c c}alves}, {Polehampton}, {Saraceno}, {Sauvage}, {Sidher},
  {Spencer}, {Swinyard}, {Ward-Thompson}, \& {White}}]{2010A&A...518L..99A}
{Anderson}, L.~D., {Zavagno}, A., {Rod{\'o}n}, J.~A., {et~al.} 2010, \aap, 518,
  L99

\bibitem[{{Andr{\'e}} {et~al.}(2010){Andr{\'e}}, {Men'shchikov}, {Bontemps},
  {K{\"o}nyves}, {Motte}, {Schneider}, {Didelon}, {Minier}, {Saraceno},
  {Ward-Thompson}, {di Francesco}, {White}, {Molinari}, {Testi}, {Abergel},
  {Griffin}, {Henning}, {Royer}, {Mer{\'{\i}}n}, {Vavrek}, {Attard},
  {Arzoumanian}, {Wilson}, {Ade}, {Aussel}, {Baluteau}, {Benedettini},
  {Bernard}, {Blommaert}, {Cambr{\'e}sy}, {Cox}, {di Giorgio}, {Hargrave},
  {Hennemann}, {Huang}, {Kirk}, {Krause}, {Launhardt}, {Leeks}, {Le Pennec},
  {Li}, {Martin}, {Maury}, {Olofsson}, {Omont}, {Peretto}, {Pezzuto}, {Prusti},
  {Roussel}, {Russeil}, {Sauvage}, {Sibthorpe}, {Sicilia-Aguilar}, {Spinoglio},
  {Waelkens}, {Woodcraft}, \& {Zavagno}}]{2010A&A...518L.102A}
{Andr{\'e}}, P., {Men'shchikov}, A., {Bontemps}, S., {et~al.} 2010, \aap, 518,
  L102

\bibitem[{{Aniano} {et~al.}(2011){Aniano}, {Draine}, {Gordon}, \&
  {Sandstrom}}]{2011PASP..123.1218A}
{Aniano}, G., {Draine}, B.~T., {Gordon}, K.~D., \& {Sandstrom}, K. 2011, \pasp,
  123, 1218

\bibitem[{{Arthur} \& {Hoare}(2006)}]{2006ApJS..165..283A}
{Arthur}, S.~J. \& {Hoare}, M.~G. 2006, \apjs, 165, 283

\bibitem[{{Arthur} {et~al.}(2004){Arthur}, {Kurtz}, {Franco}, \&
  {Albarr{\'a}n}}]{2004ApJ...608..282A}
{Arthur}, S.~J., {Kurtz}, S.~E., {Franco}, J., \& {Albarr{\'a}n}, M.~Y. 2004,
  \apj, 608, 282

\bibitem[{{Battersby} {et~al.}(2011){Battersby}, {Bally}, {Ginsburg},
  {Bernard}, {Brunt}, {Fuller}, {Martin}, {Molinari}, {Mottram}, {Peretto},
  {Testi}, \& {Thompson}}]{2011A&A...535A.128B}
{Battersby}, C., {Bally}, J., {Ginsburg}, A., {et~al.} 2011, \aap, 535, A128

\bibitem[{{Beckwith} {et~al.}(1990){Beckwith}, {Sargent}, {Chini}, \&
  {Guesten}}]{1990AJ.....99..924B}
{Beckwith}, S.~V.~W., {Sargent}, A.~I., {Chini}, R.~S., \& {Guesten}, R. 1990,
  \aj, 99, 924

\bibitem[{{Benjamin} {et~al.}(2003){Benjamin}, {Churchwell}, {Babler}, {Bania},
  {Clemens}, {Cohen}, {Dickey}, {Indebetouw}, {Jackson}, {Kobulnicky},
  {Lazarian}, {Marston}, {Mathis}, {Meade}, {Seager}, {Stolovy}, {Watson},
  {Whitney}, {Wolff}, \& {Wolfire}}]{2003PASP..115..953B}
{Benjamin}, R.~A., {Churchwell}, E., {Babler}, B.~L., {et~al.} 2003, \pasp,
  115, 953

\bibitem[{{Bessell} \& {Brett}(1988)}]{1988PASP..100.1134B}
{Bessell}, M.~S. \& {Brett}, J.~M. 1988, \pasp, 100, 1134

\bibitem[{{Borissova} {et~al.}(2011){Borissova}, {Bonatto}, {Kurtev}, {Clarke},
  {Pe{\~n}aloza}, {Sale}, {Minniti}, {Alonso-Garc{\'{\i}}a}, {Artigau},
  {Barb{\'a}}, {Bica}, {Baume}, {Catelan}, {Chen{\`e}}, {Dias}, {Folkes},
  {Froebrich}, {Geisler}, {de Grijs}, {Hanson}, {Hempel}, {Ivanov}, {Kumar},
  {Lucas}, {Mauro}, {Moni Bidin}, {Rejkuba}, {Saito}, {Tamura}, \&
  {Toledo}}]{2011A&A...532A.131B}
{Borissova}, J., {Bonatto}, C., {Kurtev}, R., {et~al.} 2011, \aap, 532, A131

\bibitem[{{Bronfman} {et~al.}(1996){Bronfman}, {Nyman}, \&
  {May}}]{1996A&AS..115...81B}
{Bronfman}, L., {Nyman}, L.-A., \& {May}, J. 1996, \aaps, 115, 81

\bibitem[{{Carey} {et~al.}(2009){Carey}, {Noriega-Crespo}, {Mizuno}, {Shenoy},
  {Paladini}, {Kraemer}, {Price}, {Flagey}, {Ryan}, {Ingalls}, {Kuchar},
  {Pinheiro Gon{\c c}alves}, {Indebetouw}, {Billot}, {Marleau}, {Padgett},
  {Rebull}, {Bressert}, {Ali}, {Molinari}, {Martin}, {Berriman}, {Boulanger},
  {Latter}, {Miville-Deschenes}, {Shipman}, \& {Testi}}]{2009PASP..121...76C}
{Carey}, S.~J., {Noriega-Crespo}, A., {Mizuno}, D.~R., {et~al.} 2009, \pasp,
  121, 76

\bibitem[{{Caswell} {et~al.}(2010){Caswell}, {Fuller}, {Green}, {Avison},
  {Breen}, {Brooks}, {Burton}, {Chrysostomou}, {Cox}, {Diamond}, {Ellingsen},
  {Gray}, {Hoare}, {Masheder}, {McClure-Griffiths}, {Pestalozzi}, {Phillips},
  {Quinn}, {Thompson}, {Voronkov}, {Walsh}, {Ward-Thompson}, {Wong-McSweeney},
  {Yates}, \& {Cohen}}]{2010MNRAS.404.1029C}
{Caswell}, J.~L., {Fuller}, G.~A., {Green}, J.~A., {et~al.} 2010, \mnras, 404,
  1029

\bibitem[{{Chavarr{\'{\i}}a} {et~al.}(2008){Chavarr{\'{\i}}a}, {Allen}, {Hora},
  {Brunt}, \& {Fazio}}]{2008ApJ...682..445C}
{Chavarr{\'{\i}}a}, L.~A., {Allen}, L.~E., {Hora}, J.~L., {Brunt}, C.~M., \&
  {Fazio}, G.~G. 2008, \apj, 682, 445

\bibitem[{{Churchwell}(2002)}]{2002ARA&A..40...27C}
{Churchwell}, E. 2002, \araa, 40, 27

\bibitem[{{Churchwell} {et~al.}(2004){Churchwell}, {Whitney}, {Babler},
  {Indebetouw}, {Meade}, {Watson}, {Wolff}, {Wolfire}, {Bania}, {Benjamin},
  {Clemens}, {Cohen}, {Devine}, {Dickey}, {Heitsch}, {Jackson}, {Kobulnicky},
  {Marston}, {Mathis}, {Mercer}, {Stauffer}, \&
  {Stolovy}}]{2004ApJS..154..322C}
{Churchwell}, E., {Whitney}, B.~A., {Babler}, B.~L., {et~al.} 2004, \apjs, 154,
  322

\bibitem[{{Das} {et~al.}(2017){Das}, {Tej}, {Vig}, {Liu}, {Liu}, {Ishwara
  Chandra}, \& {Ghosh}}]{2017MNRAS.472.4750D}
{Das}, S.~R., {Tej}, A., {Vig}, S., {et~al.} 2017, \mnras, 472, 4750

\bibitem[{{Davies} {et~al.}(2011){Davies}, {Hoare}, {Lumsden}, {Hosokawa},
  {Oudmaijer}, {Urquhart}, {Mottram}, \& {Stead}}]{2011MNRAS.416..972D}
{Davies}, B., {Hoare}, M.~G., {Lumsden}, S.~L., {et~al.} 2011, \mnras, 416, 972

\bibitem[{{Dyson} \& {Williams}(1980)}]{1980pim..book.....D}
{Dyson}, J.~E. \& {Williams}, D.~A. 1980, {Physics of the interstellar medium}

\bibitem[{{Everett} \& {Churchwell}(2010)}]{2010ApJ...713..592E}
{Everett}, J.~E. \& {Churchwell}, E. 2010, \apj, 713, 592

\bibitem[{{Faimali} {et~al.}(2012){Faimali}, {Thompson}, {Hindson}, {Urquhart},
  {Pestalozzi}, {Carey}, {Shenoy}, {Veneziani}, {Molinari}, \&
  {Clark}}]{2012MNRAS.426..402F}
{Faimali}, A., {Thompson}, M.~A., {Hindson}, L., {et~al.} 2012, \mnras, 426,
  402

\bibitem[{{Fazio} {et~al.}(2004){Fazio}, {Hora}, {Allen}, {Ashby}, {Barmby},
  {Deutsch}, {Huang}, {Kleiner}, {Marengo}, {Megeath}, {Melnick}, {Pahre},
  {Patten}, {Polizotti}, {Smith}, {Taylor}, {Wang}, {Willner}, {Hoffmann},
  {Pipher}, {Forrest}, {McMurty}, {McCreight}, {McKelvey}, {McMurray}, {Koch},
  {Moseley}, {Arendt}, {Mentzell}, {Marx}, {Losch}, {Mayman}, {Eichhorn},
  {Krebs}, {Jhabvala}, {Gezari}, {Fixsen}, {Flores}, {Shakoorzadeh}, {Jungo},
  {Hakun}, {Workman}, {Karpati}, {Kichak}, {Whitley}, {Mann}, {Tollestrup},
  {Eisenhardt}, {Stern}, {Gorjian}, {Bhattacharya}, {Carey}, {Nelson},
  {Glaccum}, {Lacy}, {Lowrance}, {Laine}, {Reach}, {Stauffer}, {Surace},
  {Wilson}, {Wright}, {Hoffman}, {Domingo}, \& {Cohen}}]{2004ApJS..154...10F}
{Fazio}, G.~G., {Hora}, J.~L., {Allen}, L.~E., {et~al.} 2004, \apjs, 154, 10

\bibitem[{{Froebrich}(2013)}]{2013IJAA....3..161F}
{Froebrich}, D. 2013, International Journal of Astronomy and Astrophysics, 3,
  161

\bibitem[{{Garay} \& {Lizano}(1999)}]{1999PASP..111.1049G}
{Garay}, G. \& {Lizano}, S. 1999, \pasp, 111, 1049

\bibitem[{{Griffin} {et~al.}(2010){Griffin}, {Abergel}, {Abreu}, {Ade},
  {Andr{\'e}}, {Augueres}, {Babbedge}, {Bae}, {Baillie}, {Baluteau}, {Barlow},
  {Bendo}, {Benielli}, {Bock}, {Bonhomme}, {Brisbin}, {Brockley-Blatt},
  {Caldwell}, {Cara}, {Castro-Rodriguez}, {Cerulli}, {Chanial}, {Chen},
  {Clark}, {Clements}, {Clerc}, {Coker}, {Communal}, {Conversi}, {Cox},
  {Crumb}, {Cunningham}, {Daly}, {Davis}, {de Antoni}, {Delderfield}, {Devin},
  {di Giorgio}, {Didschuns}, {Dohlen}, {Donati}, {Dowell}, {Dowell}, {Duband},
  {Dumaye}, {Emery}, {Ferlet}, {Ferrand}, {Fontignie}, {Fox}, {Franceschini},
  {Frerking}, {Fulton}, {Garcia}, {Gastaud}, {Gear}, {Glenn}, {Goizel},
  {Griffin}, {Grundy}, {Guest}, {Guillemet}, {Hargrave}, {Harwit}, {Hastings},
  {Hatziminaoglou}, {Herman}, {Hinde}, {Hristov}, {Huang}, {Imhof}, {Isaak},
  {Israelsson}, {Ivison}, {Jennings}, {Kiernan}, {King}, {Lange}, {Latter},
  {Laurent}, {Laurent}, {Leeks}, {Lellouch}, {Levenson}, {Li}, {Li},
  {Lilienthal}, {Lim}, {Liu}, {Lu}, {Madden}, {Mainetti}, {Marliani}, {McKay},
  {Mercier}, {Molinari}, {Morris}, {Moseley}, {Mulder}, {Mur}, {Naylor},
  {Nguyen}, {O'Halloran}, {Oliver}, {Olofsson}, {Olofsson}, {Orfei}, {Page},
  {Pain}, {Panuzzo}, {Papageorgiou}, {Parks}, {Parr-Burman}, {Pearce},
  {Pearson}, {P{\'e}rez-Fournon}, {Pinsard}, {Pisano}, {Podosek}, {Pohlen},
  {Polehampton}, {Pouliquen}, {Rigopoulou}, {Rizzo}, {Roseboom}, {Roussel},
  {Rowan-Robinson}, {Rownd}, {Saraceno}, {Sauvage}, {Savage}, {Savini},
  {Sawyer}, {Scharmberg}, {Schmitt}, {Schneider}, {Schulz}, {Schwartz},
  {Shafer}, {Shupe}, {Sibthorpe}, {Sidher}, {Smith}, {Smith}, {Smith},
  {Spencer}, {Stobie}, {Sudiwala}, {Sukhatme}, {Surace}, {Stevens}, {Swinyard},
  {Trichas}, {Tourette}, {Triou}, {Tseng}, {Tucker}, {Turner}, {Vaccari},
  {Valtchanov}, {Vigroux}, {Virique}, {Voellmer}, {Walker}, {Ward}, {Waskett},
  {Weilert}, {Wesson}, {White}, {Whitehouse}, {Wilson}, {Winter}, {Woodcraft},
  {Wright}, {Xu}, {Zavagno}, {Zemcov}, {Zhang}, \&
  {Zonca}}]{2010A&A...518L...3G}
{Griffin}, M.~J., {Abergel}, A., {Abreu}, A., {et~al.} 2010, \aap, 518, L3

\bibitem[{{Hildebrand}(1983)}]{1983QJRAS..24..267H}
{Hildebrand}, R.~H. 1983, \qjras, 24, 267

\bibitem[{{Hoare} {et~al.}(2007){Hoare}, {Kurtz}, {Lizano}, {Keto}, \&
  {Hofner}}]{2007prpl.conf..181H}
{Hoare}, M.~G., {Kurtz}, S.~E., {Lizano}, S., {Keto}, E., \& {Hofner}, P. 2007,
  Protostars and Planets V, 181

\bibitem[{{Hoare} {et~al.}(1991){Hoare}, {Roche}, \&
  {Glencross}}]{1991MNRAS.251..584H}
{Hoare}, M.~G., {Roche}, P.~F., \& {Glencross}, W.~M. 1991, \mnras, 251, 584

\bibitem[{{Hoq} {et~al.}(2013){Hoq}, {Jackson}, {Foster}, {Sanhueza},
  {Guzm{\'a}n}, {Whitaker}, {Claysmith}, {Rathborne}, {Vasyunina}, \&
  {Vasyunin}}]{2013ApJ...777..157H}
{Hoq}, S., {Jackson}, J.~M., {Foster}, J.~B., {et~al.} 2013, \apj, 777, 157

\bibitem[{{Immer} {et~al.}(2014){Immer}, {Cyganowski}, {Reid}, \&
  {Menten}}]{2014A&A...563A..39I}
{Immer}, K., {Cyganowski}, C., {Reid}, M.~J., \& {Menten}, K.~M. 2014, \aap,
  563, A39

\bibitem[{{Inoue} {et~al.}(2001){Inoue}, {Hirashita}, \&
  {Kamaya}}]{2001ApJ...555..613I}
{Inoue}, A.~K., {Hirashita}, H., \& {Kamaya}, H. 2001, \apj, 555, 613

\bibitem[{{Israel}(1978)}]{1978A&A....70..769I}
{Israel}, F.~P. 1978, \aap, 70, 769

\bibitem[{{Kauffmann} {et~al.}(2008){Kauffmann}, {Bertoldi}, {Bourke}, {Evans},
  \& {Lee}}]{2008A&A...487..993K}
{Kauffmann}, J., {Bertoldi}, F., {Bourke}, T.~L., {Evans}, II, N.~J., \& {Lee},
  C.~W. 2008, \aap, 487, 993

\bibitem[{{Kobulnicky} \& {Johnson}(1999)}]{1999ApJ...527..154K}
{Kobulnicky}, H.~A. \& {Johnson}, K.~E. 1999, \apj, 527, 154

\bibitem[{{Kwan}(1997)}]{1997ApJ...489..284K}
{Kwan}, J. 1997, \apj, 489, 284

\bibitem[{{Lada}(1987)}]{1987IAUS..115....1L}
{Lada}, C.~J. 1987, in IAU Symposium, Vol. 115, Star Forming Regions, ed.
  M.~{Peimbert} \& J.~{Jugaku}, 1--17

\bibitem[{{Lada} \& {Adams}(1992)}]{1992ApJ...393..278L}
{Lada}, C.~J. \& {Adams}, F.~C. 1992, \apj, 393, 278

\bibitem[{{Lal} \& {Rao}(2007)}]{2007MNRAS.374.1085L}
{Lal}, D.~V. \& {Rao}, A.~P. 2007, \mnras, 374, 1085

\bibitem[{{Launhardt} {et~al.}(2013){Launhardt}, {Stutz}, {Schmiedeke},
  {Henning}, {Krause}, {Balog}, {Beuther}, {Birkmann}, {Hennemann},
  {Kainulainen}, {Khanzadyan}, {Linz}, {Lippok}, {Nielbock}, {Pitann}, {Ragan},
  {Risacher}, {Schmalzl}, {Shirley}, {Stecklum}, {Steinacker}, \&
  {Tackenberg}}]{2013A&A...551A..98L}
{Launhardt}, R., {Stutz}, A.~M., {Schmiedeke}, A., {et~al.} 2013, \aap, 551,
  A98

\bibitem[{{Liu} {et~al.}(2017){Liu}, {Figueira}, {Zavagno}, {Hill},
  {Schneider}, {Men'shchikov}, {Russeil}, {Motte}, {Tig{\'e}}, {Deharveng},
  {Anderson}, {Li}, {Wu}, {Yuan}, \& {Huang}}]{2017A&A...602A..95L}
{Liu}, H.-L., {Figueira}, M., {Zavagno}, A., {et~al.} 2017, \aap, 602, A95

\bibitem[{{Liu} {et~al.}(2016){Liu}, {Li}, {Wu}, {Yuan}, {Liu}, {Dubner},
  {Paron}, {Ortega}, {Molinari}, {Huang}, {Zavagno}, {Samal}, {Huang}, \&
  {Zhang}}]{2016ApJ...818...95L}
{Liu}, H.-L., {Li}, J.-Z., {Wu}, Y., {et~al.} 2016, \apj, 818, 95

\bibitem[{{Luisi} {et~al.}(2016){Luisi}, {Anderson}, {Balser}, {Bania}, \&
  {Wenger}}]{2016ApJ...824..125L}
{Luisi}, M., {Anderson}, L.~D., {Balser}, D.~S., {Bania}, T.~M., \& {Wenger},
  T.~V. 2016, \apj, 824, 125

\bibitem[{{Lumsden} {et~al.}(2013){Lumsden}, {Hoare}, {Urquhart}, {Oudmaijer},
  {Davies}, {Mottram}, {Cooper}, \& {Moore}}]{2013ApJS..208...11L}
{Lumsden}, S.~L., {Hoare}, M.~G., {Urquhart}, J.~S., {et~al.} 2013, \apjs, 208,
  11

\bibitem[{{Mac Low} {et~al.}(1991){Mac Low}, {van Buren}, {Wood}, \&
  {Churchwell}}]{1991ApJ...369..395M}
{Mac Low}, M.-M., {van Buren}, D., {Wood}, D.~O.~S., \& {Churchwell}, E. 1991,
  \apj, 369, 395

\bibitem[{{Mallick} {et~al.}(2015){Mallick}, {Ojha}, {Tamura}, {Linz}, {Samal},
  \& {Ghosh}}]{2015MNRAS.447.2307M}
{Mallick}, K.~K., {Ojha}, D.~K., {Tamura}, M., {et~al.} 2015, \mnras, 447, 2307

\bibitem[{{Mart{\'{\i}}n-Hern{\'a}ndez}
  {et~al.}(2003){Mart{\'{\i}}n-Hern{\'a}ndez}, {Bik}, {Kaper}, {Tielens}, \&
  {Hanson}}]{2003A&A...405..175M}
{Mart{\'{\i}}n-Hern{\'a}ndez}, N.~L., {Bik}, A., {Kaper}, L., {Tielens},
  A.~G.~G.~M., \& {Hanson}, M.~M. 2003, \aap, 405, 175

\bibitem[{{Martins} {et~al.}(2005){Martins}, {Schaerer}, \&
  {Hillier}}]{2005A&A...436.1049M}
{Martins}, F., {Schaerer}, D., \& {Hillier}, D.~J. 2005, \aap, 436, 1049

\bibitem[{{Meyer} {et~al.}(1997){Meyer}, {Calvet}, \&
  {Hillenbrand}}]{1997AJ....114..288M}
{Meyer}, M.~R., {Calvet}, N., \& {Hillenbrand}, L.~A. 1997, \aj, 114, 288

\bibitem[{{Mezger} \& {Henderson}(1967)}]{1967ApJ...147..471M}
{Mezger}, P.~G. \& {Henderson}, A.~P. 1967, \apj, 147, 471

\bibitem[{{Minniti} {et~al.}(2010){Minniti}, {Lucas}, {Emerson}, {Saito},
  {Hempel}, {Pietrukowicz}, {Ahumada}, {Alonso}, {Alonso-Garcia}, {Arias},
  {Bandyopadhyay}, {Barb{\'a}}, {Barbuy}, {Bedin}, {Bica}, {Borissova},
  {Bronfman}, {Carraro}, {Catelan}, {Clari{\'a}}, {Cross}, {de Grijs},
  {D{\'e}k{\'a}ny}, {Drew}, {Fari{\~n}a}, {Feinstein}, {Fern{\'a}ndez
  Laj{\'u}s}, {Gamen}, {Geisler}, {Gieren}, {Goldman}, {Gonzalez}, {Gunthardt},
  {Gurovich}, {Hambly}, {Irwin}, {Ivanov}, {Jord{\'a}n}, {Kerins}, {Kinemuchi},
  {Kurtev}, {L{\'o}pez-Corredoira}, {Maccarone}, {Masetti}, {Merlo},
  {Messineo}, {Mirabel}, {Monaco}, {Morelli}, {Padilla}, {Palma}, {Parisi},
  {Pignata}, {Rejkuba}, {Roman-Lopes}, {Sale}, {Schreiber}, {Schr{\"o}der},
  {Smith}, {}, {Soto}, {Tamura}, {Tappert}, {Thompson}, {Toledo}, {Zoccali}, \&
  {Pietrzynski}}]{2010NewA...15..433M}
{Minniti}, D., {Lucas}, P.~W., {Emerson}, J.~P., {et~al.} 2010, \na, 15, 433

\bibitem[{{Molinari} {et~al.}(2010){Molinari}, {Swinyard}, {Bally}, {Barlow},
  {Bernard}, {Martin}, {Moore}, {Noriega-Crespo}, {Plume}, {Testi}, {Zavagno},
  {Abergel}, {Ali}, {Anderson}, {Andr{\'e}}, {Baluteau}, {Battersby},
  {Beltr{\'a}n}, {Benedettini}, {Billot}, {Blommaert}, {Bontemps}, {Boulanger},
  {Brand}, {Brunt}, {Burton}, {Calzoletti}, {Carey}, {Caselli}, {Cesaroni},
  {Cernicharo}, {Chakrabarti}, {Chrysostomou}, {Cohen}, {Compiegne}, {de
  Bernardis}, {de Gasperis}, {di Giorgio}, {Elia}, {Faustini}, {Flagey},
  {Fukui}, {Fuller}, {Ganga}, {Garcia-Lario}, {Glenn}, {Goldsmith}, {Griffin},
  {Hoare}, {Huang}, {Ikhenaode}, {Joblin}, {Joncas}, {Juvela}, {Kirk},
  {Lagache}, {Li}, {Lim}, {Lord}, {Marengo}, {Marshall}, {Masi}, {Massi},
  {Matsuura}, {Minier}, {Miville-Desch{\^e}nes}, {Montier}, {Morgan}, {Motte},
  {Mottram}, {M{\"u}ller}, {Natoli}, {Neves}, {Olmi}, {Paladini}, {Paradis},
  {Parsons}, {Peretto}, {Pestalozzi}, {Pezzuto}, {Piacentini}, {Piazzo},
  {Polychroni}, {Pomar{\`e}s}, {Popescu}, {Reach}, {Ristorcelli}, {Robitaille},
  {Robitaille}, {Rod{\'o}n}, {Roy}, {Royer}, {Russeil}, {Saraceno}, {Sauvage},
  {Schilke}, {Schisano}, {Schneider}, {Schuller}, {Schulz}, {Sibthorpe},
  {Smith}, {Smith}, {Spinoglio}, {Stamatellos}, {Strafella}, {Stringfellow},
  {Sturm}, {Taylor}, {Thompson}, {Traficante}, {Tuffs}, {Umana}, {Valenziano},
  {Vavrek}, {Veneziani}, {Viti}, {Waelkens}, {Ward-Thompson}, {White},
  {Wilcock}, {Wyrowski}, {Yorke}, \& {Zhang}}]{2010A&A...518L.100M}
{Molinari}, S., {Swinyard}, B., {Bally}, J., {et~al.} 2010, \aap, 518, L100

\bibitem[{{Morales} {et~al.}(2013){Morales}, {Wyrowski}, {Schuller}, \&
  {Menten}}]{2013A&A...560A..76M}
{Morales}, E.~F.~E., {Wyrowski}, F., {Schuller}, F., \& {Menten}, K.~M. 2013,
  \aap, 560, A76

\bibitem[{{Mottram} {et~al.}(2011){Mottram}, {Hoare}, {Davies}, {Lumsden},
  {Oudmaijer}, {Urquhart}, {Moore}, {Cooper}, \& {Stead}}]{2011ApJ...730L..33M}
{Mottram}, J.~C., {Hoare}, M.~G., {Davies}, B., {et~al.} 2011, \apjl, 730, L33

\bibitem[{{M{\"u}cke} {et~al.}(2002){M{\"u}cke}, {Koribalski}, {Moffat},
  {Corcoran}, \& {Stevens}}]{2002ApJ...571..366M}
{M{\"u}cke}, A., {Koribalski}, B.~S., {Moffat}, A.~F.~J., {Corcoran}, M.~F., \&
  {Stevens}, I.~R. 2002, \apj, 571, 366

\bibitem[{{Nandakumar} {et~al.}(2016){Nandakumar}, {Veena}, {Vig}, {Tej},
  {Ghosh}, \& {Ojha}}]{2016AJ....152..146N}
{Nandakumar}, G., {Veena}, V.~S., {Vig}, S., {et~al.} 2016, \aj, 152, 146

\bibitem[{{Ojha} {et~al.}(2004{\natexlab{a}}){Ojha}, {Tamura}, {Nakajima},
  {Fukagawa}, {Sugitani}, {Nagashima}, {Nagayama}, {Nagata}, {Sato}, {Pickles},
  \& {Ogura}}]{2004ApJ...608..797O}
{Ojha}, D.~K., {Tamura}, M., {Nakajima}, Y., {et~al.} 2004{\natexlab{a}}, \apj,
  608, 797

\bibitem[{{Ojha} {et~al.}(2004{\natexlab{b}}){Ojha}, {Tamura}, {Nakajima},
  {Fukagawa}, {Sugitani}, {Nagashima}, {Nagayama}, {Nagata}, {Sato}, {Vig},
  {Ghosh}, {Pickles}, {Momose}, \& {Ogura}}]{2004ApJ...616.1042O}
{Ojha}, D.~K., {Tamura}, M., {Nakajima}, Y., {et~al.} 2004{\natexlab{b}}, \apj,
  616, 1042

\bibitem[{{Osterbrock}(1989)}]{1989agna.book.....O}
{Osterbrock}, D.~E. 1989, {Astrophysics of gaseous nebulae and active galactic
  nuclei}

\bibitem[{{Paladini} {et~al.}(2012){Paladini}, {Umana}, {Veneziani},
  {Noriega-Crespo}, {Anderson}, {Piacentini}, {Pinheiro Gon{\c c}alves},
  {Paradis}, {Tibbs}, {Bernard}, \& {Natoli}}]{2012ApJ...760..149P}
{Paladini}, R., {Umana}, G., {Veneziani}, M., {et~al.} 2012, \apj, 760, 149

\bibitem[{{Panagia}(1973)}]{1973AJ.....78..929P}
{Panagia}, N. 1973, \aj, 78, 929

\bibitem[{{Paron} {et~al.}(2011){Paron}, {Petriella}, \&
  {Ortega}}]{2011A&A...525A.132P}
{Paron}, S., {Petriella}, A., \& {Ortega}, M.~E. 2011, \aap, 525, A132

\bibitem[{{Peretto} {et~al.}(2010){Peretto}, {Fuller}, {Plume}, {Anderson},
  {Bally}, {Battersby}, {Beltran}, {Bernard}, {Calzoletti}, {Digiorgio},
  {Faustini}, {Kirk}, {Lenfestey}, {Marshall}, {Martin}, {Molinari}, {Montier},
  {Motte}, {Ristorcelli}, {Rod{\'o}n}, {Smith}, {Traficante}, {Veneziani},
  {Ward-Thompson}, \& {Wilcock}}]{2010A&A...518L..98P}
{Peretto}, N., {Fuller}, G.~A., {Plume}, R., {et~al.} 2010, \aap, 518, L98

\bibitem[{{Poglitsch} {et~al.}(2010){Poglitsch}, {Waelkens}, {Geis},
  {Feuchtgruber}, {Vandenbussche}, {Rodriguez}, {Krause}, {Renotte}, {van
  Hoof}, {Saraceno}, {Cepa}, {Kerschbaum}, {Agn{\`e}se}, {Ali}, {Altieri},
  {Andreani}, {Augueres}, {Balog}, {Barl}, {Bauer}, {Belbachir}, {Benedettini},
  {Billot}, {Boulade}, {Bischof}, {Blommaert}, {Callut}, {Cara}, {Cerulli},
  {Cesarsky}, {Contursi}, {Creten}, {De Meester}, {Doublier}, {Doumayrou},
  {Duband}, {Exter}, {Genzel}, {Gillis}, {Gr{\"o}zinger}, {Henning},
  {Herreros}, {Huygen}, {Inguscio}, {Jakob}, {Jamar}, {Jean}, {de Jong},
  {Katterloher}, {Kiss}, {Klaas}, {Lemke}, {Lutz}, {Madden}, {Marquet},
  {Martignac}, {Mazy}, {Merken}, {Montfort}, {Morbidelli}, {M{\"u}ller},
  {Nielbock}, {Okumura}, {Orfei}, {Ottensamer}, {Pezzuto}, {Popesso},
  {Putzeys}, {Regibo}, {Reveret}, {Royer}, {Sauvage}, {Schreiber}, {Stegmaier},
  {Schmitt}, {Schubert}, {Sturm}, {Thiel}, {Tofani}, {Vavrek}, {Wetzstein},
  {Wieprecht}, \& {Wiezorrek}}]{2010A&A...518L...2P}
{Poglitsch}, A., {Waelkens}, C., {Geis}, N., {et~al.} 2010, \aap, 518, L2

\bibitem[{{Quireza} {et~al.}(2006){Quireza}, {Rood}, {Bania}, {Balser}, \&
  {Maciel}}]{2006ApJ...653.1226Q}
{Quireza}, C., {Rood}, R.~T., {Bania}, T.~M., {Balser}, D.~S., \& {Maciel},
  W.~J. 2006, \apj, 653, 1226

\bibitem[{{Ranjan Das} {et~al.}(2016){Ranjan Das}, {Tej}, {Vig}, {Ghosh}, \&
  {Ishwara Chandra}}]{2016AJ....152..152R}
{Ranjan Das}, S., {Tej}, A., {Vig}, S., {Ghosh}, S.~K., \& {Ishwara Chandra},
  C.~H. 2016, \aj, 152, 152

\bibitem[{{Reid} \& {Ho}(1985)}]{1985ApJ...288L..17R}
{Reid}, M.~J. \& {Ho}, P.~T.~P. 1985, \apjl, 288, L17

\bibitem[{{Rieke} \& {Lebofsky}(1985)}]{1985ApJ...288..618R}
{Rieke}, G.~H. \& {Lebofsky}, M.~J. 1985, \apj, 288, 618

\bibitem[{{Rodriguez} {et~al.}(1993){Rodriguez}, {Marti}, {Canto}, {Moran}, \&
  {Curiel}}]{1993RMxAA..25...23R}
{Rodriguez}, L.~F., {Marti}, J., {Canto}, J., {Moran}, J.~M., \& {Curiel}, S.
  1993, \rmxaa, 25, 23

\bibitem[{{Rosero} {et~al.}(2016){Rosero}, {Hofner}, {Claussen}, {Kurtz},
  {Cesaroni}, {Araya}, {Carrasco-Gonz{\'a}lez}, {Rodr{\'{\i}}guez}, {Menten},
  {Wyrowski}, {Loinard}, \& {Ellingsen}}]{2016ApJS..227...25R}
{Rosero}, V., {Hofner}, P., {Claussen}, M., {et~al.} 2016, \apjs, 227, 25

\bibitem[{{Roth} {et~al.}(2014){Roth}, {Stahler}, \&
  {Keto}}]{2014MNRAS.438.1335R}
{Roth}, N., {Stahler}, S.~W., \& {Keto}, E. 2014, \mnras, 438, 1335

\bibitem[{{Rubin}(1968)}]{1968ApJ...153..761R}
{Rubin}, R.~H. 1968, \apj, 153, 761

\bibitem[{{Russeil} {et~al.}(2016){Russeil}, {Tig{\'e}}, {Adami}, {Anderson},
  {Schneider}, {Zavagno}, {Samal}, {Amram}, {Guennou}, {Le Coarer}, {Walsh},
  {Longmore}, \& {Purcell}}]{2016A&A...587A.135R}
{Russeil}, D., {Tig{\'e}}, J., {Adami}, C., {et~al.} 2016, \aap, 587, A135

\bibitem[{{S{\'a}nchez-Monge} {et~al.}(2013){S{\'a}nchez-Monge}, {Kurtz},
  {Palau}, {Estalella}, {Shepherd}, {Lizano}, {Franco}, \&
  {Garay}}]{2013ApJ...766..114S}
{S{\'a}nchez-Monge}, {\'A}., {Kurtz}, S., {Palau}, A., {et~al.} 2013, \apj,
  766, 114

\bibitem[{{Schmiedeke} {et~al.}(2016){Schmiedeke}, {Schilke}, {M{\"o}ller},
  {S{\'a}nchez-Monge}, {Bergin}, {Comito}, {Csengeri}, {Lis}, {Molinari},
  {Qin}, \& {Rolffs}}]{2016A&A...588A.143S}
{Schmiedeke}, A., {Schilke}, P., {M{\"o}ller}, T., {et~al.} 2016, \aap, 588,
  A143

\bibitem[{{Schraml} \& {Mezger}(1969)}]{1969ApJ...156..269S}
{Schraml}, J. \& {Mezger}, P.~G. 1969, \apj, 156, 269

\bibitem[{{Schuller} {et~al.}(2009){Schuller}, {Menten}, {Contreras},
  {Wyrowski}, {Schilke}, {Bronfman}, {Henning}, {Walmsley}, {Beuther},
  {Bontemps}, {Cesaroni}, {Deharveng}, {Garay}, {Herpin}, {Lefloch}, {Linz},
  {Mardones}, {Minier}, {Molinari}, {Motte}, {Nyman}, {Reveret}, {Risacher},
  {Russeil}, {Schneider}, {Testi}, {Troost}, {Vasyunina}, {Wienen}, {Zavagno},
  {Kovacs}, {Kreysa}, {Siringo}, \& {Wei{\ss}}}]{2009A&A...504..415S}
{Schuller}, F., {Menten}, K.~M., {Contreras}, Y., {et~al.} 2009, \aap, 504, 415

\bibitem[{{Skrutskie} {et~al.}(2006){Skrutskie}, {Cutri}, {Stiening},
  {Weinberg}, {Schneider}, {Carpenter}, {Beichman}, {Capps}, {Chester},
  {Elias}, {Huchra}, {Liebert}, {Lonsdale}, {Monet}, {Price}, {Seitzer},
  {Jarrett}, {Kirkpatrick}, {Gizis}, {Howard}, {Evans}, {Fowler}, {Fullmer},
  {Hurt}, {Light}, {Kopan}, {Marsh}, {McCallon}, {Tam}, {Van Dyk}, \&
  {Wheelock}}]{2006AJ....131.1163S}
{Skrutskie}, M.~F., {Cutri}, R.~M., {Stiening}, R., {et~al.} 2006, \aj, 131,
  1163

\bibitem[{{Spitzer}(1978)}]{1978ppim.book.....S}
{Spitzer}, L. 1978, {Physical processes in the interstellar medium}

\bibitem[{{Sugitani} {et~al.}(2002){Sugitani}, {Tamura}, {Nakajima},
  {Nagashima}, {Nagayama}, {Nakaya}, {Pickles}, {Nagata}, {Sato}, {Fukuda}, \&
  {Ogura}}]{2002ApJ...565L..25S}
{Sugitani}, K., {Tamura}, M., {Nakajima}, Y., {et~al.} 2002, \apjl, 565, L25

\bibitem[{{Swarup} {et~al.}(1991){Swarup}, {Ananthakrishnan}, {Kapahi}, {Rao},
  {Subrahmanya}, \& {Kulkarni}}]{1991CuSc...60...95S}
{Swarup}, G., {Ananthakrishnan}, S., {Kapahi}, V.~K., {et~al.} 1991, Current
  Science, Vol.~60, NO.2/JAN25, P.~95, 1991, 60, 95

\bibitem[{{Tej} {et~al.}(2006){Tej}, {Ojha}, {Ghosh}, {Kulkarni}, {Verma},
  {Vig}, \& {Prabhu}}]{2006A&A...452..203T}
{Tej}, A., {Ojha}, D.~K., {Ghosh}, S.~K., {et~al.} 2006, \aap, 452, 203

\bibitem[{{Tenorio-Tagle}(1979)}]{1979A&A....71...59T}
{Tenorio-Tagle}, G. 1979, \aap, 71, 59

\bibitem[{{Urquhart} {et~al.}(2007{\natexlab{a}}){Urquhart}, {Busfield},
  {Hoare}, {Lumsden}, {Clarke}, {Moore}, {Mottram}, \&
  {Oudmaijer}}]{2007A&A...461...11U}
{Urquhart}, J.~S., {Busfield}, A.~L., {Hoare}, M.~G., {et~al.}
  2007{\natexlab{a}}, \aap, 461, 11

\bibitem[{{Urquhart} {et~al.}(2007{\natexlab{b}}){Urquhart}, {Busfield},
  {Hoare}, {Lumsden}, {Oudmaijer}, {Moore}, {Gibb}, {Purcell}, {Burton}, \&
  {Marechal}}]{2007A&A...474..891U}
{Urquhart}, J.~S., {Busfield}, A.~L., {Hoare}, M.~G., {et~al.}
  2007{\natexlab{b}}, \aap, 474, 891

\bibitem[{{Urquhart} {et~al.}(2014){Urquhart}, {Figura}, {Moore}, {Hoare},
  {Lumsden}, {Mottram}, {Thompson}, \& {Oudmaijer}}]{2014MNRAS.437.1791U}
{Urquhart}, J.~S., {Figura}, C.~C., {Moore}, T.~J.~T., {et~al.} 2014, \mnras,
  437, 1791

\bibitem[{{van Buren} \& {Mac Low}(1992)}]{1992ApJ...394..534V}
{van Buren}, D. \& {Mac Low}, M.-M. 1992, \apj, 394, 534

\bibitem[{{van Buren} {et~al.}(1990){van Buren}, {Mac Low}, {Wood}, \&
  {Churchwell}}]{1990ApJ...353..570V}
{van Buren}, D., {Mac Low}, M.-M., {Wood}, D.~O.~S., \& {Churchwell}, E. 1990,
  \apj, 353, 570

\bibitem[{{van der Walt} {et~al.}(1995){van der Walt}, {Gaylard}, \&
  {MacLeod}}]{1995A&AS..110...81V}
{van der Walt}, D.~J., {Gaylard}, M.~J., \& {MacLeod}, G.~C. 1995, \aaps, 110,
  81

\bibitem[{{Veena} {et~al.}(2016){Veena}, {Vig}, {Tej}, {Varricatt}, {Ghosh},
  {Chandrasekhar}, \& {Ashok}}]{2016MNRAS.456.2425V}
{Veena}, V.~S., {Vig}, S., {Tej}, A., {et~al.} 2016, \mnras, 456, 2425

\bibitem[{{Vig} {et~al.}(2007){Vig}, {Ghosh}, {Ojha}, \&
  {Verma}}]{2007A&A...463..175V}
{Vig}, S., {Ghosh}, S.~K., {Ojha}, D.~K., \& {Verma}, R.~P. 2007, \aap, 463,
  175

\bibitem[{{Watson} {et~al.}(2008){Watson}, {Povich}, {Churchwell}, {Babler},
  {Chunev}, {Hoare}, {Indebetouw}, {Meade}, {Robitaille}, \&
  {Whitney}}]{2008ApJ...681.1341W}
{Watson}, C., {Povich}, M.~S., {Churchwell}, E.~B., {et~al.} 2008, \apj, 681,
  1341

\bibitem[{{Wenger} {et~al.}(2013){Wenger}, {Bania}, {Balser}, \&
  {Anderson}}]{2013ApJ...764...34W}
{Wenger}, T.~V., {Bania}, T.~M., {Balser}, D.~S., \& {Anderson}, L.~D. 2013,
  \apj, 764, 34

\bibitem[{{Williams} {et~al.}(1994){Williams}, {de Geus}, \&
  {Blitz}}]{1994ApJ...428..693W}
{Williams}, J.~P., {de Geus}, E.~J., \& {Blitz}, L. 1994, \apj, 428, 693

\bibitem[{{Wood} \& {Churchwell}(1989)}]{1989ApJS...69..831W}
{Wood}, D.~O.~S. \& {Churchwell}, E. 1989, \apjs, 69, 831

\bibitem[{{Yu} {et~al.}(2015){Yu}, {Wang}, \& {Li}}]{2015MNRAS.446.2566Y}
{Yu}, N., {Wang}, J.-J., \& {Li}, N. 2015, \mnras, 446, 2566

\bibitem[{{Zhang} \& {Wang}(2012)}]{2012A&A...544A..11Z}
{Zhang}, C.~P. \& {Wang}, J.~J. 2012, \aap, 544, A11

\bibitem[{{Zhu} {et~al.}(2005){Zhu}, {Lacy}, {Jaffe}, {Richter}, \&
  {Greathouse}}]{2005ApJ...631..381Z}
{Zhu}, Q.-F., {Lacy}, J.~H., {Jaffe}, D.~T., {Richter}, M.~J., \& {Greathouse},
  T.~K. 2005, \apj, 631, 381

\bibitem[{{Zhu} {et~al.}(2008){Zhu}, {Lacy}, {Jaffe}, {Richter}, \&
  {Greathouse}}]{2008ApJS..177..584Z}
{Zhu}, Q.-F., {Lacy}, J.~H., {Jaffe}, D.~T., {Richter}, M.~J., \& {Greathouse},
  T.~K. 2008, \apjs, 177, 584

\bibitem[{{Zoonematkermani} {et~al.}(1990){Zoonematkermani}, {Helfand},
  {Becker}, {White}, \& {Perley}}]{1990ApJS...74..181Z}
{Zoonematkermani}, S., {Helfand}, D.~J., {Becker}, R.~H., {White}, R.~L., \&
  {Perley}, R.~A. 1990, \apjs, 74, 181

\end{thebibliography}


\end{document}